\documentclass{jaa}
\usepackage{natbib}
\usepackage[dvipsnames]{xcolor}
\definecolor{xlinkcolor}{cmyk}{1,1,0,0}

 \usepackage[bookmarks=false,         
     pdfnewwindow=true,      
     colorlinks=true,    
     linkcolor=xlinkcolor,     
     citecolor=xlinkcolor,     
     filecolor=xlinkcolor,  
     urlcolor=xlinkcolor,      
final=true
 ]{hyperref}

\usepackage{comment}
\usepackage{gensymb}
\usepackage{graphicx}
\usepackage{lastpage}
\usepackage{multicol}
\usepackage{aas_macros}
\usepackage{subcaption}
\usepackage{rotating}
\usepackage{xspace}
\usepackage{multirow}
\usepackage{booktabs}
\usepackage{bm}

\newcommand{\asat}{{\em AstroSat}\xspace}
\newcommand{\fermi}{{\em Fermi}\xspace}
\newcommand{\swift}{{\em Swift}\xspace}
\newcommand{\kw}{{\em {Konus}-Wind}\xspace}

\newcommand{\mm}{Mass Model}

\newcommand{\polar}{\emph{POLAR}\xspace}
\newcommand{\integral}{\emph{INTEGRAL}\xspace}

\hypersetup{colorlinks=true,allcolors={cyan}}

\begin{document}\sloppy

\title{Investigating Polarization characteristics of GRB~200503A and GRB~201009A}

\author{\href{https://orcid.org/0000-0001-6332-1723}{Divita Saraogi}\textsuperscript{1,*},
\href{https://orcid.org/0000-0002-6657-9022}{Suman Bala}\textsuperscript{1,2,*}, 
\href{https://orcid.org/0009-0009-7042-5817}{Jitendra Joshi}\textsuperscript{3},
\href{https://orcid.org/0000-0002-2525-3464}{Shabnam Iyyani}\textsuperscript{4,5},
\href{https://orcid.org/0000-0002-6112-7609}{Varun Bhalerao}\textsuperscript{1},
\href{https://orcid.org/0009-0004-2350-7936}{J~Venkata Aditya}\textsuperscript{6},
\href{https://orcid.org/0000-0002-2208-2196}{D.S. Svinkin}\textsuperscript{7},
\href{https://orcid.org/0000-0002-1153-6340}{D.D. Frederiks}\textsuperscript{7},
\href{https://orcid.org/0000-0002-3942-8341}{A.L. Lysenko}\textsuperscript{7},
\href{https://orcid.org/0000-0001-9477-5437}{A.V. Ridnaia}\textsuperscript{7},
\href{https://orcid.org/0000-0002-5868-0868 }{A.S. Kozyrev}\textsuperscript{8},
\href{}{D.V. Golovin}\textsuperscript{8},
\href{https://orcid.org/0000-0001-7204-2101}{I.G. Mitrofanov}\textsuperscript{8},
\href{}{M.L. Litvak}\textsuperscript{8},
\href{}{A.B. Sanin}\textsuperscript{8},
\href{https://orcid.org/0000-0001-9856-1866}{Tanmoy Chattopadyay}\textsuperscript{9},
\href{https://orcid.org/0000-0001-6621-259X}{Soumya Gupta}\textsuperscript{10},
\href{https://orcid.org/0000-0003-3630-9440}{Gaurav Waratkar}\textsuperscript{1},
\href{https://orcid.org/0000-0003-3352-3142}{Dipankar Bhattacharya}\textsuperscript{11},
\href{https://orcid.org/0000-0002-2050-0913}{Santosh Vadawale}\textsuperscript{12},
\href{https://orcid.org/0000-0003-1589-2075}{Gulab Dewangan}\textsuperscript{3}
}

\affilOne{\textsuperscript{1} Department of Physics, Indian Institute of Technology Bombay, Mumbai, Maharashtra 400076, India\\}
\affilTwo{\textsuperscript{2} Science and Technology Institute, Universities Space Research Association, Huntsville, AL 35805, USA\\}
\affilThree{\textsuperscript{3} Inter University Centre for Astronomy and Astrophysics, Pune, Maharashtra 411007, India\\}
\affilFour{\textsuperscript{4} School of Physics, Indian Institute of Science Education and Research, Thiruvananthapuram, 695551, India\\}
\affilFive{\textsuperscript{5} Centre of High Performance Computing, Indian Institute of Science Education and Research, Thiruvananthapuram, 695551, India\\}
\affilSix{\textsuperscript{6} Department of Computer Science and Engineering, Indian Institute of Technology Bombay, Powai, Mumbai, Maharashtra 400076, India\\}
\affilSeven{\textsuperscript{7} Ioffe Institute, Polytekhnicheskaya, 26, St. Petersburg, 194021, Russia\\}
\affilEight{\textsuperscript{8} Space Research Institute (IKI), 84/32 Profsoyuznaya, Moscow, 117997, Russia\\}
\affilNine{\textsuperscript{9} Kavli Institute of Particle Astrophysics and Cosmology, 452 Lomita Mall, Stanford, CA 94305, USA\\}
\affilTen{\textsuperscript{10} Astrophysical Science Division, Homi Bhabha National Institute, Anushakti Nagar, Mumbai, Maharashtra 400094, India\\}
\affilEleven{\textsuperscript{11} Department of Physics, Ashoka University, Sonepat, Haryana 131029, India \\}
\affilTwelve{\textsuperscript{12} Physical Research Laboratory, Ahmedabad, Gujarat 380009, India\\}

\twocolumn[{

\maketitle

\corres{divitadsaraogi@gmail.com, sbala@usra.edu}

\msinfo{---}{---}

\begin{abstract}
We present results of a comprehensive analysis of the polarization characteristics of GRB~200503A and GRB~201009A observed with the Cadmium Zinc Telluride Imager (CZTI) on board \asat. Despite these GRBs being reasonably bright, they were missed by several spacecraft and had thus far not been localized well, hindering polarization analysis. We present positions of these bursts obtained from the Inter-Planetary Network (IPN) and the newly developed CZTI localization pipeline. We then undertook polarization analyses using the standard CZTI pipeline. We cannot constrain the polarization properties for GRB~200503A, but find that GRB~201009A has a high degree of polarization. 
\end{abstract}


}]


\doinum{12.3456/s78910-011-012-3}
\artcitid{\#\#\#\#}
\volnum{000}
\year{0000}
\pgrange{1--}
\setcounter{page}{1}
\lp{\pageref{LastPage}}

\section{Introduction}\label{sec:intro}

Gamma-ray bursts (GRBs)  are the most energetic transient phenomena in the Universe \citep{discovery,1999PhR...314..575P}. They are believed to be powered by newly formed black holes \citep{woosley93, iwamoto98, macfadyen99} or magnetars \citep{eichler89, narayan92}, which result from the core collapse of massive stars \citep{woosley93, iwamoto98, macfadyen99} or the merger of two compact objects, such as binary neutron stars (BNS) or a neutron star-black hole (NS-BH) pair \citep{eichler89, narayan92, Abbott_170817A}. Despite decades of research, the prompt emission phase of GRBs—the initial, brief, and intense gamma-ray flash--remains to be fully understood. Key questions regarding the central engine, jet structure, magnetic fields, and emission mechanisms within the relativistic jet are still unresolved \citep{kumar15, zhang19}. Competing models exist that propose different mechanisms for radiation, energy dissipation, and the resulting non-thermal spectra observed during the prompt emission phase \citep{Gill2021}. These spectra are typically modeled using phenomenological fits, such as a power law with an exponential cutoff or the Band function \citep{band1993batse, gruber2014fermi}. The spectral and temporal properties of the prompt emission are consistent with multiple models, including the synchrotron model (both ordered and random), the Compton drag model, and the Photospheric model \citep{iyyani15, Zhang2016}. Polarization observations are crucial for distinguishing between these competing models, as they predict different ranges of polarization fractions for GRBs based on the source geometry \citep{toma08, covino16, mcconnell16, Gill_etal_2020, Gill2021}.

Measurement of high-energy polarization is a photon-intensive task, especially challenging by the brief and non-recurring nature of the prompt emission phase of GRBs \citep{2021JApA...42..106C,2021JApA...42...82C}. To date, more than 5000 GRBs have been observed, but polarization measurements have only been achieved for approximately 40 of them. Most of these measurements have been obtained in recent years by instruments such as \asat/CZTI~\citep{chattopadhyay19,Chattopadhyay2022} and \polar~\citep{zhang19,Kole2020}. Other missions, including \emph{BATSE}~\citep{Willis2005}, \emph{GAP}~\citep{2012ApJ...758L...1Y}, and \integral/SPI~\citep{Gotz09,Gotz13}, have also contributed a few measurements.

Different missions have reported varying levels of polarization in observed GRBs. For instance, \emph{GAP} and \emph{INTEGRAL} have measured high polarization fractions (greater than 60\%) in their sources, while \polar has observed lower polarization levels in its data \citep{2020A&A...644A.124K}. \asat has detected both high and low polarization in its sample data \citep{Chattopadhyay2022,Chand_2019}.

In a very limited number of cases, variations in polarization have been reported. For example, in GRB 100826A, a flip in the polarization angle between two pulses was observed \citep{yonetoku11,2012ApJ...758L...1Y}. GRB~170114A exhibited temporal evolution of polarization within a single pulse \citep{zhang19,Burgess_etal_2019}. A detailed study of GRB 160821A, observed with \asat, revealed a ~90$^\circ$ flip in the polarization angle, occurring twice within a single broad emission pulse \citep{Sharma_etal_2019}. Furthermore, variable prompt emission polarization was detected in GRB 171010A with \asat-CZTI \citep{Chand_2019}.

All of these observations suggest that the prompt emission phase of GRBs is highly dynamic. For a comprehensive overview of GRB polarization measurements, see \citet{mcconnell16}, \citet{2021JApA...42..106C}, \citet{Chattopadhyay2022}, and \cite{Gill2021}.

This paper presents the polarization results of two bright GRBs, GRB~200503A and GRB~201009A, using \asat CZTI. Analyzing GRB data for polarization requires knowledge of the location of the source in the sky. The localization for these GRBs had not previously been published. We obtained the localization using multiple satellites in IPN. In addition, we also used the \citet{saraogi_2024} pipeline to localize the bursts with the CZTI data. The resultant localizations from these methods are consistent, but still uncertain. Since the position of the source is an important parameter in the polarization calculation, we also examine for the first time the sensitivity of the polarization measurements to the source position in \asat~CZTI.


In Section \ref{sec:info}, we present an overview of the two GRBs under investigation, detailing their localization by the Interplanetary Network (IPN) and \asat \mm, along with their spectral characteristics. Sections \ref{sec:200503A} and \ref{sec:201009A} delve into the time-integrated and time-resolved polarization analysis for GRB~200503A and GRB~201009A respectively. In Section \ref{sec:robust}, we examine how the localization impacts our polarization results. Lastly, in Section \ref{sec:conclude}, we interpret the findings, providing scientific insights into the observations and potential emission mechanisms of these GRBs.
\section{Detection and properties of GRB~200503A and GRB~201009A}\label{sec:info}
\asat~CZTI recorded the occurrence of a long GRB~200503A on May 3, 2020, at 05:02:01.5~UT within the 40--200~keV range \citep{2020GCN.27680....1G}. The peak count rate reached $2021\pm 50$ counts/s, with a total of $28736\pm52$ counts across all four quadrants. Our refined analysis estimates that the duration of the burst $\mathrm{T_{90}}=45$~s where $\mathrm{T_{90}}$ is the time duration during which 90\% of the total detected photon counts from a GRB are recorded. It is found that 2104 Compton events are associated with this burst. This GRB was also detected by the CsI anti-coincidence (Veto) detector in the 100--500~keV energy range. Additionally, the \textit{AGILE} satellite observed the event at 05:01:44~UT, capturing a multi-peaked time profile lasting approximately 25 s \citep{2020GCN.27685....1U}.

Similarly, \asat CZTI detected the long, bright GRB~201009A on October 9, 2020, at 03:08:21.5~UT. The peak count rate reached $4215\pm69$ counts/s, with a total of $45310\pm77$ counts. The CsI anti-coincidence detector (Veto) also clearly detected the event within the 100–500~keV energy range \citep{2020GCN.28589....1G}. Our refined analysis measured $\mathrm{T_{90}}$ of 35 seconds, with 3277 Compton events associated with the burst. Additionally, \textit{AGILE} observed this burst a few seconds after emerging from the South Atlantic Anomaly (SAA) at 03:08:26 UT on October 9, 2020 \citep{2020GCN.28588....1U}. \kw has also observed these bursts but not reported them.

During the burst episodes, \fermi was in the SAA region in the case of GRB~200503A, and the source position of GRB~201009A was behind the Earth. So, \fermi could not observe any of these GRBs. These GRBs were also not detected by the \textit{ Neil Gehrels Swift Observatory}.

Spectral information is crucial for both the localization and polarization analysis of GRBs. The spectral parameters for both GRBs were obtained from the \kw data. For GRB~200503A and GRB~201009A, the spectral parameters obtained by fitting Band function to the time integrated spectra are mentioned in Table \ref{tab:konus_para}. This spectral information was used in all of our analyses.

\begin{table*}[htb!]
    \centering
    \begin{tabular}{ccccc}
        \toprule
        GRB & $\alpha$ & $\beta$ & $\text{E}_{\text{peak}}$ & Fluence   \\
        & & & (keV) & ($\text{erg} \, \text{cm}^{-2}$) \\
        \toprule
        200503A & $-1.0^{+0.08}_{-0.07} $ & $-2.61^{+0.18}_{-0.25}$ & $272^{+20.0}_{-20.0}$ & $1.3^{+0.10}_{-0.09} \times 10^{-4}$ \\
        \\
        201009A & $-0.91^{+0.05}_{-0.04}$ & $-2.20^{+0.06}_{-0.07}$ & $369^{+26}_{-25}$ & $3.74^{+0.14}_{-0.14} \times 10^{-4}$ \\
        \bottomrule
    \end{tabular}
    \caption{Band spectra for GRB~200503A and GRB~201009A, obtained from the \kw data. The fluence is for 20~keV–10~MeV energy range.}
    \label{tab:konus_para}
\end{table*}

\subsection{Localization of GRBs}\label{sec:loc}
Both of these bursts were initially not localized by any single instrument. In this work, we address this issue in two ways: first, by applying the \citet{saraogi_2024} method to localize the bursts using \asat CZTI data and second, by using joint timing information from multiple satellites via the Interplanetary Network (IPN) to create a localization map.  IPN localizations were derived
using the following data: For GRB~200503A data from \kw, \asat(CZTI), \swift(BAT), \textit{Mars-Odyssey}(HEND) were utilized;  and for GRB~201009A data from \kw, \asat(CZTI), \textit{INTEGRAL} (SPI-ACS) were employed. The final results for GRB~200503A and GRB~201009A are shown in Figure \ref{fig:IPN+Astrosat}, where the colored contours indicate different probability regions as observed by \asat. The white line overlay on the plot represents the localization region provided by the IPN.

Although this GRB was observed by these satellites, it was not reported in the Global Coordinate Network (GCN). The triangulation performed by IPN defines a broad region, depicted as the white arc in Figure \ref{fig:GRB200503A_map}, where the GRB could be located anywhere along this arc. . The white annulus shown in Figure \ref{fig:GRB201009A_map} represents the localization region for GRB~201009A as determined by IPN. 
\end{comment}

\subsubsection{GRB~200503A:}
\begin{figure}
     \centering
     \subfloat[GRB~200503A]{\includegraphics[width=1.0\linewidth]{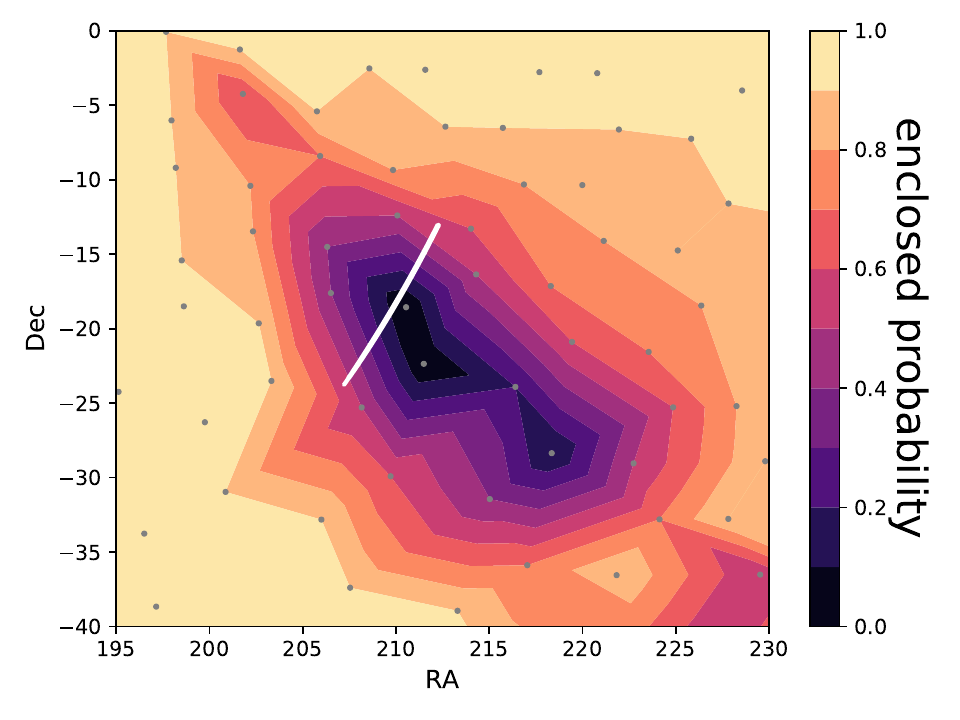}\label{fig:IPN_map_a}}\\
     \subfloat[GRB~201009A]{\includegraphics[width=1.0\linewidth]{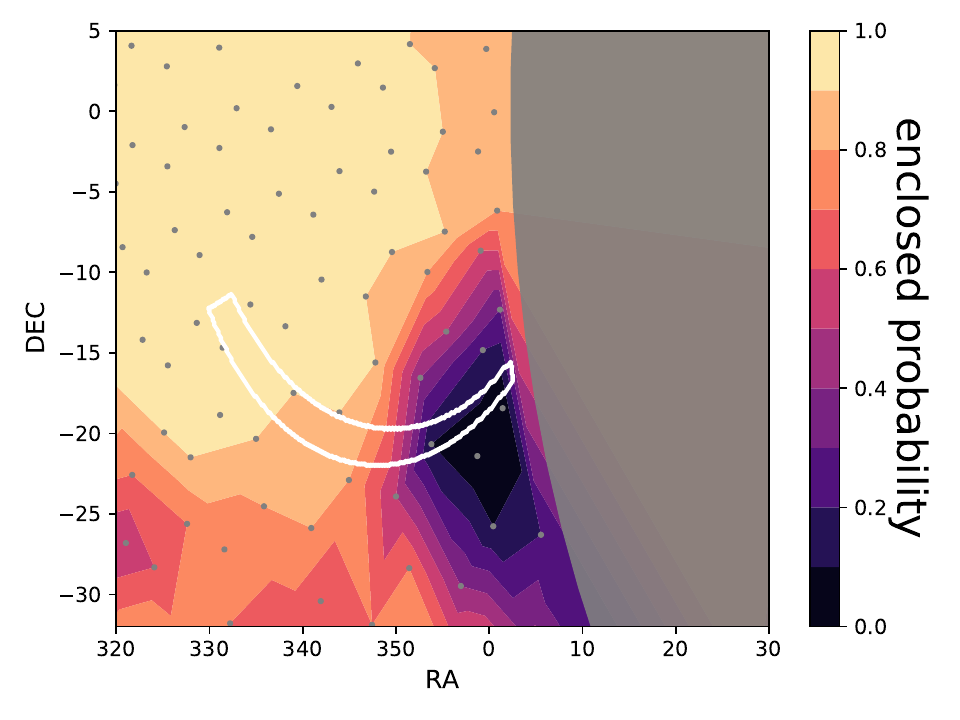}\label{fig:IPN_map_b}}
     \caption{All sky probability contour plots for GRB~200503A and GRB~201009A in RA, Dec. The white over-plotted line is the localisation map by IPN. The grey dots represent the hierarchical triangular mesh (HTM) grid points. In Figure (b) for GRB~201009A, the grey region is the region occulted by the Earth.}
     \label{fig:IPN+Astrosat}
\end{figure}

We used the spectral parameters from \kw, and the CZTI data in the 70--200 keV energy range to obtain localization  using the \asat~\mm. Our best-fit localization for GRB~200503A is at RA = $210^\circ$, Dec = $-18^\circ$ with $\theta = 38^\circ$ and $\phi = 166^\circ$. We note that this point is very close to the narrow IPN arc, as seen in Figure \ref{fig:IPN_map_a}. Given the GRB's total count of 12536 during the event interval, the $50\%$ probability area $\mathrm{A_{50}}$ according to \asat localization is 548 square degrees. 

\subsubsection{GRB~201009A:}
For the localization of GRB~201009A, we used the CZTI data in the energy range 70--200 keV and the spectral parameters from \kw. For this GRB, the CZTI localization pipeline gives a position that is just at the edge of the earth-occulted region, away from the IPN annulus. Hence, we take the best position as RA=355$\degree$, Dec =-20$\degree$ corresponding to $\theta=58 \degree$, $\phi=155\degree$ in the CZTI plane based on both localizations. As this GRB has a total of 17300 counts for the time interval of the GRB, the 50$\%$ probability area $\mathrm{A_{50}}$ according to our formula is 397 square degrees. The CZTI localization $\chi^{2}$ probability contours and the over-plotted IPN map are shown in Figure \ref{fig:IPN_map_b}.

\section{GRB~200503A}\label{sec:200503A}
\begin{figure}[htb!]
	\centering
	\includegraphics[width=0.5\textwidth]{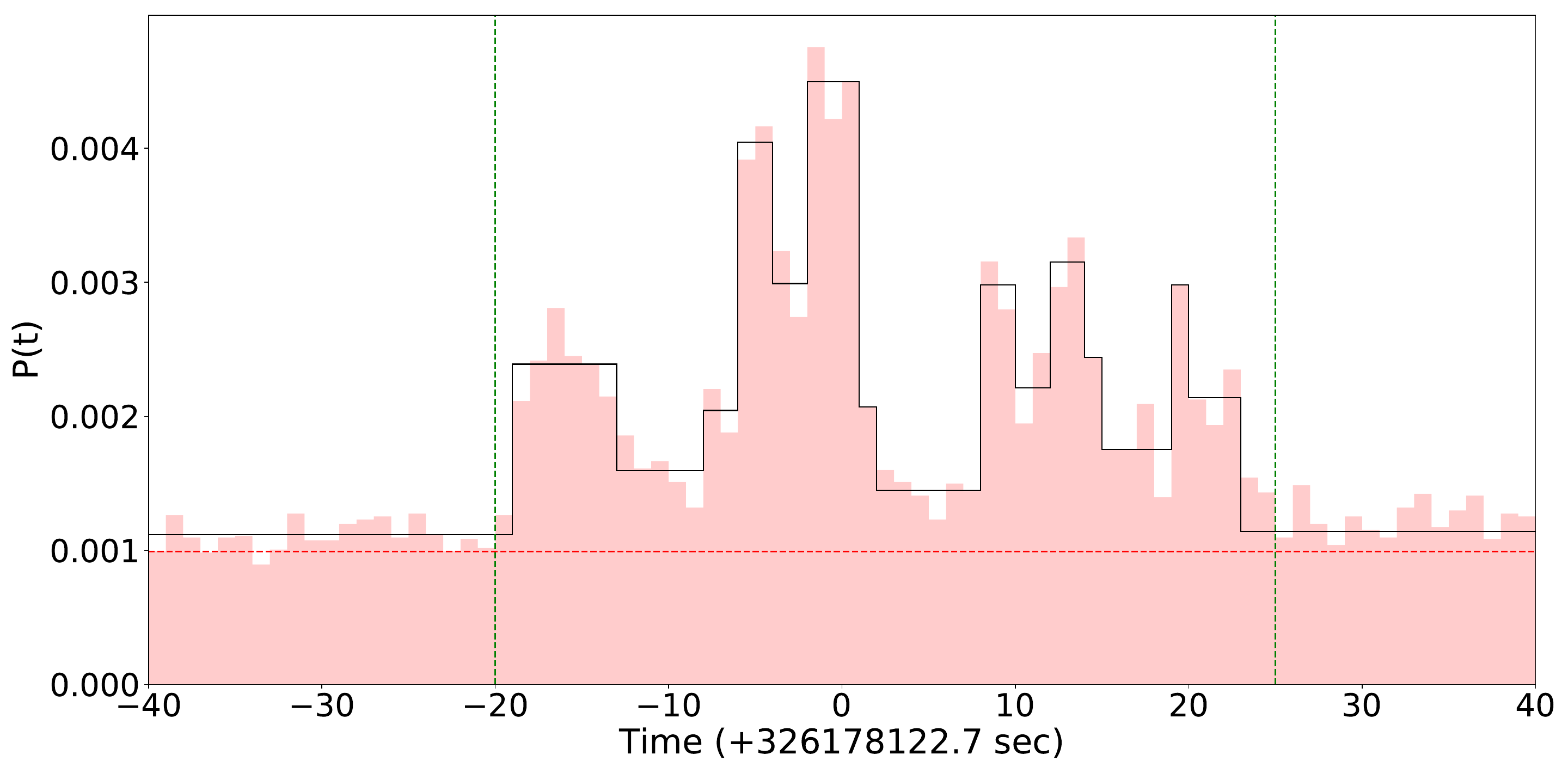}
	\caption{Bayesian block decomposed light curve for GRB~200503A. The green dashed lines indicate the start and end of the GRB. } 
 \label{fig:200503a_bay}
 \end{figure}

\begin{figure}[htb!]
	\centering
	\includegraphics[width=0.5\textwidth]{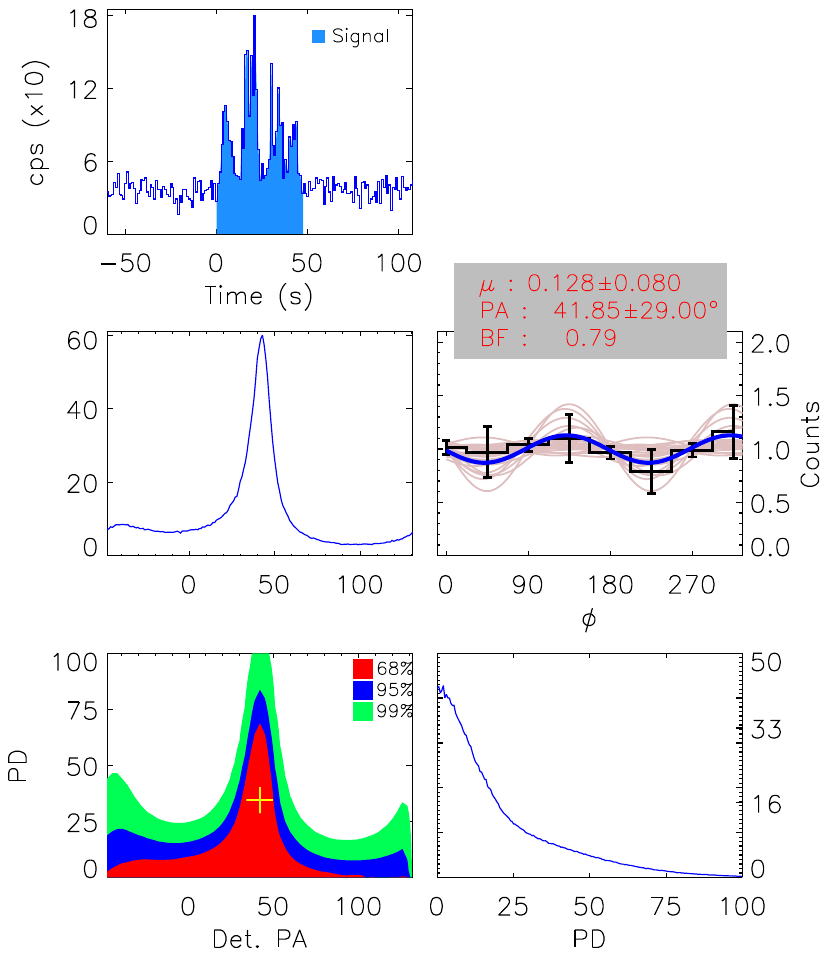}
	\caption{Time-average polarisation analysis outcome for GRB~200503A. The top figure shows a Compton light curve in the 100--600 keV range with 1~s binning, with Compton events for polarisation analysis extracted from the shaded blue region. Middle left: Posterior probability distribution for the polarisation angle derived from Markov Chain Monte Carlo (MCMC) iterations. Middle right: The modulation curve (black line and error bar) and sinusoidal fit (solid blue line), accompanied by 100 random MCMC iterations. Bottom left: Contour plot for polarisation angle and degree, indicating confidence levels of 68\%, 95\%, and 99\%. Bottom right: Posterior probability distribution for polarisation degree from MCMC iterations.}
 \label{fig:GRB200503a_integrated}
 \end{figure}

GRB 200503A is a long-duration GRB with $T_{90} = 45$~s. The Bayesian block decomposition of its light curve is shown in Figure \ref{fig:200503a_bay}. The time-integrated polarization analysis in the specified $\theta$ and $\phi$, based on our localization method, is presented in Figure \ref{fig:GRB200503a_integrated}. Our analysis results in a modulation factor of $\mu = 0.12 \pm 0.06$, a polarization angle measured in the detector plane (PA) of $46 \pm 33^\circ$, and a Bayes factor of 0.75. The low Bayes factor $(\le 2)$ and high uncertainty in the modulation factor both indicate that our data cannot constrain the polarization of GRB~200503A. 

\subsection{Time resolved spectral analysis of GRB~200503A}
The long GRB~200503A can be divided into three distinct time windows based on the Bayesian decomposition of its light curve, as shown in Figure \ref{fig:200503a_bay}. Inspired by previous works showing changes of polarization between pulses, we attempt to analyze the individual pulses. However, we found that pulse 1 had insufficient Compton counts for polarization analysis. Hence, we combined pulse 1,2 into a single time window from 0 to 25~s, and considered the third pulse independently, from t =20~s to 25~s. So now pulse 1+2 combined has 1379 Compton counts, and pulse 3 has 809 Compton counts. We find that the GRB polarization remains unconstrained in both these windows, similar to the time-averaged case.


\begin{table*}[htb!]
     \centering
     \begin{tabular}{ccccccc}
     
     \multicolumn{7}{c}{\textbf{GRB~200503A}} \\
     	\toprule
        \textbf{Intervals}  &  \textbf{Duration}  &  \textbf{Compton counts} &  \bm{$\mu$ } & \textbf{ PA} & \textbf{ PF} &  \textbf{Bayes factor}\\
        \toprule
        Total duration & 45 & 2104 & $ 0.12\pm0.07$    &  --   & $\le 33.66$  &  0.79 \\
       
\bottomrule
\multicolumn{7}{c}{ } \\
\multicolumn{7}{c}{\textbf{GRB~201009A}} \\

     	\toprule
        \textbf{Region}  &  \textbf{Duration}  &  \textbf{Compton counts} &  \bm{$\mu$}  &  \textbf{PA} &  \textbf{PF }&  \textbf{Bayes factor}\\
        \toprule
        Total duration & 35 & 3277 &  $0.18\pm0.06$  & $36\pm6$ & $78\pm30$ & 7.26 \\
        Peak 1 & 20 & 2493 &  $0.14\pm0.06$          & $31\pm10$ & $71\pm31$ & 5.14\\
        Peak 2 & 15 & 816 &  $0.48\pm 0.41$          & -- & Unconstrained & 1.44\\
\bottomrule
     \end{tabular}
     \caption{Compton counts and polarisation parameters for GRB~200503A and GRB~201009A.}
     \label{tab:time_resolved_201009}
 \end{table*}

\section{GRB~201009A}\label{sec:201009A}
\begin{figure}[htb!]
	\centering
	\includegraphics[width=0.5\textwidth]{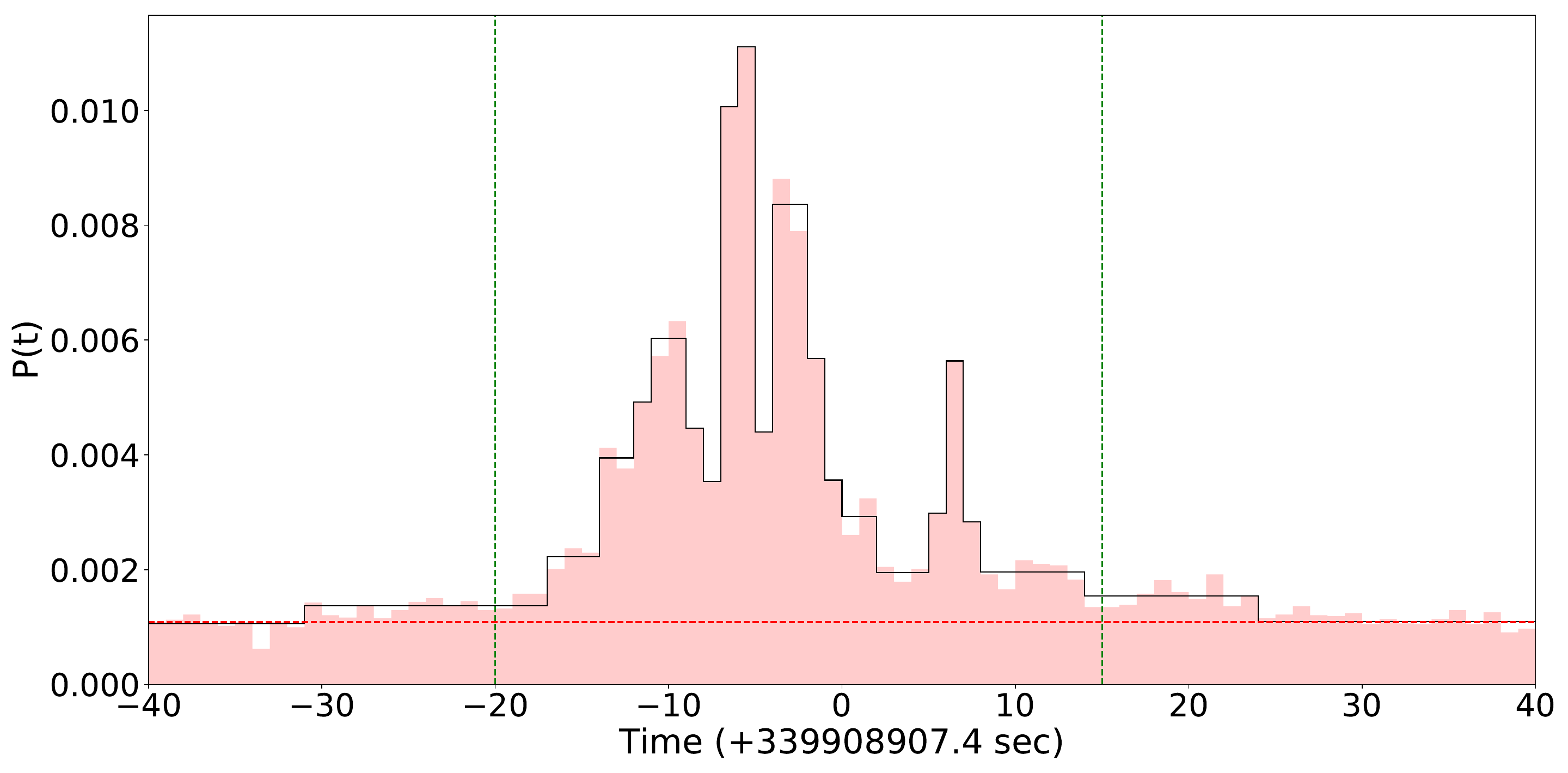}
	\caption{Bayesian block decomposed light curve for GRB~201009A. The green dashed lines indicate the start and end of the GRB.} 
 \label{fig:201009a_bay}
 \end{figure}

CZTI detected GRB~201009A with 3277 background subtracted Compton counts in a total duration of 35~s. We performed a polarization analysis on the complete burst (as described in \cite{Chattopadhyay2022}. The intervals between bursts were determined using the Bayesian block algorithm \citep{scargle1998studies,Scargle2013,Burgess2014}. The Bayesian block decomposed light curve is shown in Figure \ref{fig:201009a_bay}. 

 \begin{figure}[htb!]
	\centering
	\includegraphics[width=0.5\textwidth]{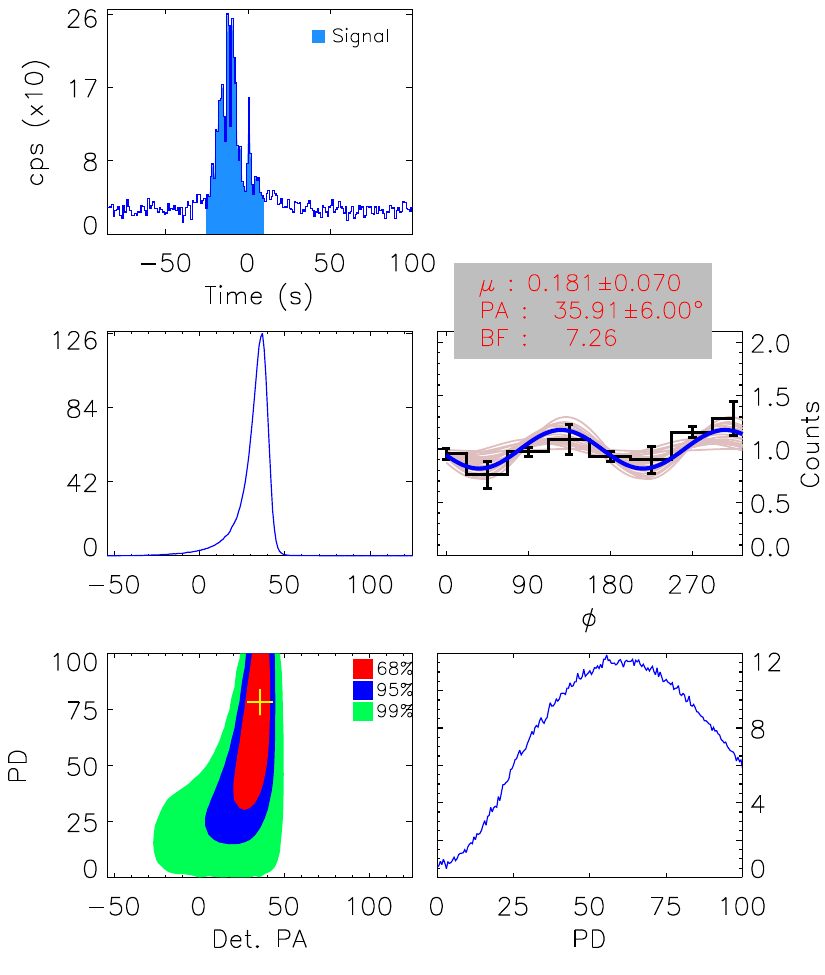}
	\caption{Same as Figure \ref{fig:GRB200503a_integrated} but for GRB~201009A.}
 \label{fig:200503a_int}
 \end{figure} 

We conducted a time-integrated polarization analysis resulting in polarization fraction of 78\% and polarization angle of $35\pm 6^\circ$. This result is statistically significant with a Bayes factor of 7.26.
\subsection{Time resolved polarization analysis}
The long GRB~201009A lightcurve shows two regions: one main peak region followed by a second peak region. We conducted a polarization analysis for each region of these two regions. In the first region, corresponding to a 20~s time window, we recorded 2493 Compton counts. The time-resolved analysis reveals that the first peak is highly polarized, with a polarization fraction of 70\% and a polarization angle of $30^\circ$. This finding is highly significant, with a Bayes factor of 5.15, indicating that the signal is genuinely polarized and that the likelihood of unpolarized radiation producing such sinusoidal modulation in the azimuthal angle distribution is low. 

\begin{figure*}
    \begin{subfigure}[t]{0.5\textwidth}
        \includegraphics[width=\textwidth]{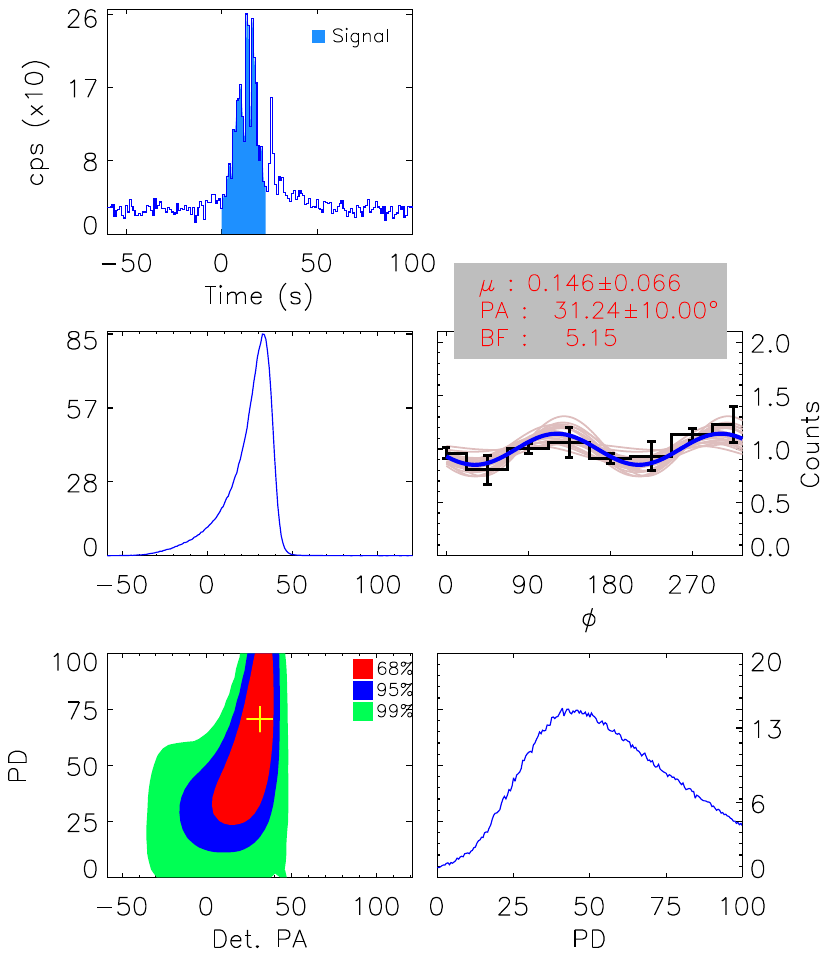}
    \end{subfigure}
    \begin{subfigure}[t]{0.5\textwidth}
        \includegraphics[width=\textwidth]{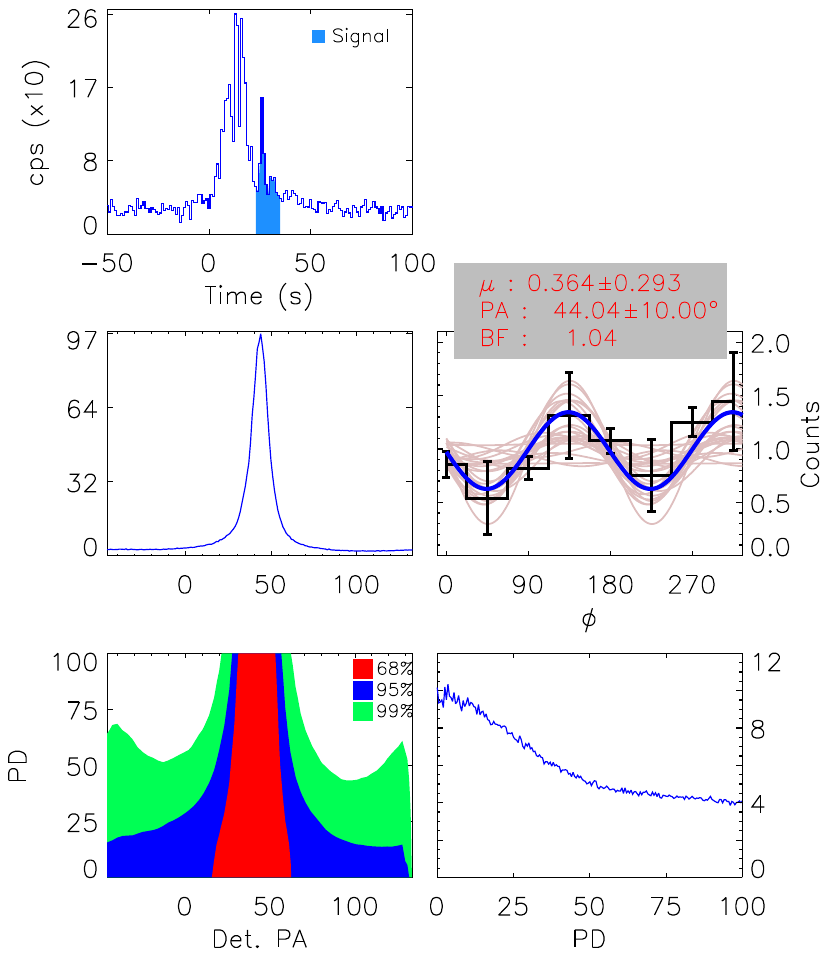}
    \end{subfigure}
    \caption{ Same as Figure \ref{fig:200503a_int}, but for time resolved cases. Left: for the time bin 0 to 20s. Right: for time bin 20 s to 35 s.}
    \label{fig:GRB201009A_time_resolved}
\end{figure*}

For the second time window, we analyzed a 15~s interval derived from our Bayesian block analysis. However, during this period, the number of Compton counts fell below the threshold required to meet the Compton criteria. As illustrated in Figure \ref{fig:GRB201009A_time_resolved}, the results for this interval exhibit high uncertainty. The primary contribution to the polarization of this burst is thus attributed to the first peak. Table \ref{tab:time_resolved_201009} gives the time resolved parameters for GRB~201009A and Figure \ref{fig:GRB201009A_time_resolved} shows the time resolved analysis.

\section{Sensitivity to localization uncertainties}\label{sec:robust}
The analysis of polarization in CZTI relies completely on simulations undertaken with the \asat~\mm~as discussed in \citet{Chattopadhyay2022}. A key input for these simulations is the position of the burst in the sky. Since there were non-negligible uncertainties in the locations of both our bursts, we repeated our analyses by considering various source locations along the IPN arcs. We focused our analysis on the sky region within the 50\% containment area ($\mathrm{A_{50}}$) as determined by CZTI. Additionally, we restricted our analysis to a 5-degree radius around the estimated location due to the computationally intensive nature of GRB polarization simulations.

The time-integrated results for GRB~200503A do not exhibit any polarization signatures, as discussed in Section \ref{sec:200503A}, a trend that persists across the six selected positions. Table \ref{tab:robust} provides the PA and PF values for GRB~200503A. Here, we still give the PA and PF values only to show that similar results and low Bayes factors were obtained in all the cases.

In our examination of GRB~201009A, we identified consistent polarization within a radius of 5 degrees. The polarization parameters measured at six distinct positions for this gamma-ray burst are presented in Table \ref{tab:robust}. The values of PA and PF for all six studied positions are consistent within error bars, and we get an acceptably high Bayes factor in each case.

  \begin{table*}[htb!]
     \centering
     \begin{tabular}{cccccccc}
     \multicolumn{8}{c}{\textbf{GRB~200503A}} \\
    \toprule
          RA &  Dec &  $\theta$ &  $\phi$  & $\mu$ &  PA &  PF &    Bayes factor\\
        \toprule
211 &  -18 &  37.8 &  167.7  &    $0.12\pm0.08$&      $45\pm29$ &  $31\pm21$&  0.76\\
207 &  -25 &  41.5 &  155.9  &    $0.13\pm0.08$&      $45\pm24$ &  $35\pm21$&  0.79\\
208 &  -23 &  40.5 &  159.8 &     $0.13\pm0.09$&      $45\pm22$ &  $37\pm26$&  0.80\\
209 &  -20 &  38.4 &  163.5 &     $0.12\pm0.07$&      $46\pm28$ &  $32\pm12$&  0.77\\
210 &  -17 &  36.4 &  167.4 &     $0.12\pm0.08$&      $46\pm25$ &  $46\pm22$&  0.79\\
212 &  -15 &  36.0 &  172.1 &     $0.12\pm0.07$&      $45\pm27$ &  $32\pm20$&  0.77\\
\bottomrule
\multicolumn{8}{c}{ } \\
 \multicolumn{8}{c}{\textbf{GRB~201009A}} \\
\toprule
          RA &  Dec &  $\theta$ &  $\phi$  & $\mu$ &  PA &  PF &    Bayes factor\\
        \toprule
 357.0&	-20.2&  59.7&    155.5&    	$0.18\pm0.07$&   $37\pm6$&    $76\pm30$&    5.81\\
 354.7&   -21.0&  57.6&	 154.6&	    $0.18\pm0.07$&	 $37\pm5$&    $73\pm28$&	 6.09\\
 353.6&   -21.3&  56.6&	 154.2&		$0.17\pm0.07$&   $37\pm6$&    $75\pm30$&	 4.55\\
 352.5&   -21.5&	55.5&	 153.9&	    $0.18\pm0.07$&	 $38\pm5$&    $71\pm28$&	 5.42\\
 351.3&	-21.7&	54.5&	 153.6&	    $0.18\pm0.07$&	 $37\pm5$&	  $79\pm31$&	 5.26\\
 350.2&	-21.7&	53.4&	 153.4&	    $0.18\pm0.07$&	 $37\pm5$&    $66\pm26$&    6.09\\
\bottomrule
     \end{tabular}
     \caption{Coordinates for each position at which polarization analysis was conducted, along with the associated polarization results for GRB~2000503A and GRB~201009A. Note that for GRB~200503A the net polarisation is unconstrained as discussed in section \ref{sec:200503A}. Here, we give nominal best-fit values despite the high error bars only to highlight the comparison.}
     \label{tab:robust}
 \end{table*}

\section{Conclusions and Discussions}\label{sec:conclude}
Investigation into the polarization of GRB prompt emission remains a relatively unexplored area in astronomical observations. This task is notably challenging due to the transient nature of GRBs, leading to limited photon statistics, as well as the complex instrument systematics involved. Additionally, the lack of localization information poses challenges for polarization studies by instruments without localization capability. We presented the first joint localization and polarization analysis of two GRBs using CZTI. Our examination indicates that the GRB localization results derived from the \asat~Mass Model align with the localization map provided by the IPN.

Previous polarimetric studies of GRBs observed by \asat CZTI and POLAR suggests that the time-integrated GRB emissions generally exhibit lower polarization levels \citep{Chattopadhyay2022,Kole2020}. This can be attributed to two possible scenarios, assuming that the bursts are viewed on the axis:
(a) The burst might be intrinsically polarized, but variations in the polarization angle across multiple emission pulses can result in an overall lower polarization. This scenario has been demonstrated in detailed polarimetric studies of GRBs such as GRB 160821A \citep{Sharma_etal_2019} and GRB 170114A \citep{Burgess_etal_2019}, where the underlying emission pulses could be due to synchrotron emission in magnetic fields ordered at least on scales of $1/\Gamma$.
(b) The burst might be intrinsically unpolarized. In this case, the underlying observed non-thermal spectral radiation could be linked to synchrotron emission originating in random magnetic fields or Compton drag \citep{toma2009statistical,Gill_etal_2020}. Given that the polarimetric study focuses on the energy range around the burst $E_{peak}$, the lack of polarization could also support subphotospheric emission models, where the burst peak is expected to be unpolarized due to its formation deep within the burst outflow at high optical depths \citep{Lundman_etal_2018}.

GRB~200503A is a multi-pulsed GRB. Our analysis cannot constrain the polarization of GRB~200503A. The GRB shows three main pulses - we combined them into 2 episodes to get enough Compton counts and found that the polarization for individual episodes are also unconstrained.  

In contrast, GRB~201009A, despite having multi-pulsed emission, exhibits a high degree of polarization in its time-integrated emission, with high statistical significance (Bayes Factor $>5$). Bayesian binning identifies two main emission episodes. Time-resolved polarimetry shows that the primary emission episode, occurring from 0 to 20 seconds, is highly polarized, while the subsequent emission from 20 to 35 seconds cannot be constrained. Given the low brightness of the second episode, we conclude that the polarization behavior of the primary episode significantly influences the time-integrated analysis (Figure \ref{fig:GRB201009A_time_resolved}). Considering the burst is viewed on-axis, the high polarization fraction (PF $> 25\%$, Figure \ref{fig:GRB201009A_time_resolved}) observed throughout the burst duration suggests that the underlying radiation is likely to be synchrotron radiation produced in an ordered transverse magnetic field \citep{toma2009statistical,Gill_etal_2020}.


Furthermore, our assessment of the stability of polarization results across different locations shows only minor fluctuations in polarization parameters, with the modulation 
factor ($\mu$), polarization angle (PA), and Bayes factor remaining nearly constant. We can conclude that up to 5$\degree$ radius, there are no significant changes in the polarization parameters with respect to the position. Enhancements in polarization data analysis through a unified framework that enables simultaneous fitting of both spectrum and polarization from various instruments will produce unbiased, high-quality results. Attempts to undertake joint location-spectral-polarisation analysis are under way, and will be presented elsewhere.
\section*{Acknowledgements}
This publication draws on data from the \asat mission, carried out by the Indian Space Research Organization (ISRO) and archived at the Indian Space Science Data Centre (ISSDC). The CZT-Imager instrument represents a collaborative effort from several Indian institutes, including the Tata Institute of Fundamental Research in Mumbai, the Vikram Sarabhai Space Centre in Thiruvananthapuram, the ISRO Satellite Centre in Bengaluru, the Inter University Centre for Astronomy and Astrophysics in Pune, the Physical Research Laboratory in Ahmedabad, and the Space Application Centre in Ahmedabad. We express our gratitude to the extensive technical teams at these institutes for their significant contributions. This work benefited greatly from the IPN data, for which we sincerely thank late Prof. Kevin Hurley.

We also acknowledge the use of the Pegasus HPC at the Inter University Centre for Astronomy and Astrophysics (IUCAA), Pune. This research utilized various software tools, including Python, AstroPy \citep{astropy}, NumPy \citep{numpy}, Matplotlib \citep{matplotlib}, IDL Astrolib \citep{landsman93}, FTOOLS \citep{blackburn95}, C, and C++. S.I. is supported by DST INSPIRE Faculty Scheme (IFA19-PH245), SERB SRG Grant (SRG/2022/000211). 

\vspace{-1em}


\bibliography{main_biblio}

\begin{thebibliography}{}
\expandafter\ifx\csname natexlab\endcsname\relax\def\natexlab#1{#1}\fi

\bibitem[{Abbott {$et~al$.}(2017)Abbott, Abbott, Abbott, Acernese, Ackley, Adams, Adams, Addesso, Adhikari, Adya, {$et~al$.}}]{Abbott_170817A}
Abbott, B.~P., Abbott, R., Abbott, T., {$et~al$.} 2017, \apjl, 848, L13

\bibitem[{Band {$et~al$.}(1993)Band, Matteson, Ford, Schaefer, Palmer, Teegarden, Cline, Briggs, Paciesas, Pendleton, {$et~al$.}}]{band1993batse}
Band, D., Matteson, J., Ford, L., {$et~al$.} 1993, The Astrophysical Journal, 413, 281

\bibitem[{Blackburn(1995)}]{blackburn95}
Blackburn, J.~K. 1995, Astronomical Data Analysis Software and Systems IV, 77

\bibitem[{{Burgess} {$et~al$.}(2019){Burgess}, {Kole}, {Berlato}, {Greiner}, {Vianello}, {Produit}, {Li}, \& {Sun}}]{Burgess_etal_2019}
{Burgess}, J.~M., {Kole}, M., {Berlato}, F., {$et~al$.} 2019, \aap, 627, A105

\bibitem[{{Burgess} {$et~al$.}(2014){Burgess}, {Preece}, {Connaughton}, {Briggs}, {Goldstein}, {Bhat}, {Greiner}, {Gruber}, {Kienlin}, {Kouveliotou}, {McGlynn}, {Meegan}, {Paciesas}, {Rau}, {Xiong}, {Axelsson}, {Baring}, {Dermer}, {Iyyani}, {Kocevski}, {Omodei}, {Ryde}, \& {Vianello}}]{Burgess2014}
{Burgess}, J.~M., {Preece}, R.~D., {Connaughton}, V., {$et~al$.} 2014, \apj, 784, 17

\bibitem[{Chand {$et~al$.}(2019)Chand, Chattopadhyay, Oganesyan, Rao, Vadawale, Bhattacharya, Bhalerao, \& Misra}]{Chand_2019}
Chand, V., Chattopadhyay, T., Oganesyan, G., {$et~al$.} 2019, The Astrophysical Journal, 874, 70

\bibitem[{{Chattopadhyay}(2021)}]{2021JApA...42..106C}
{Chattopadhyay}, T. 2021, Journal of Astrophysics and Astronomy, 42, 106

\bibitem[{{Chattopadhyay} {$et~al$.}(2019){Chattopadhyay}, {Vadawale}, {Aarthy}, {Mithun}, {Chand}, {Ratheesh}, {Basak}, {Rao}, {Bhalerao}, {Mate}, {Arvind}, {Sharma}, \& {Bhattacharya}}]{chattopadhyay19}
{Chattopadhyay}, T., {Vadawale}, S.~V., {Aarthy}, E., {$et~al$.} 2019, \apj, 884, 123

\bibitem[{{Chattopadhyay} {$et~al$.}(2021){Chattopadhyay}, {Gupta}, {Sharma}, {Iyyani}, {Ratheesh}, {Mithun}, {Aarthy}, {Palit}, {Kumar}, {Vadawale}, {Rao}, {Bhalerao}, \& {Bhattacharya}}]{2021JApA...42...82C}
{Chattopadhyay}, T., {Gupta}, S., {Sharma}, V., {$et~al$.} 2021, Journal of Astrophysics and Astronomy, 42, 82

\bibitem[{{Chattopadhyay} {$et~al$.}(2022){Chattopadhyay}, {Gupta}, {Iyyani}, {Saraogi}, {Sharma}, {Tsvetkova}, {Ratheesh}, {Gupta}, {Mithun}, {Vaishnava}, {Prasad}, {Aarthy}, {Kumar}, {Rao}, {Vadawale}, {Bhalerao}, {Bhattacharya}, {Vibhute}, \& {Frederiks}}]{Chattopadhyay2022}
{Chattopadhyay}, T., {Gupta}, S., {Iyyani}, S., {$et~al$.} 2022, \apj, 936, 12

\bibitem[{{Covino} \& {Gotz}(2016)}]{covino16}
{Covino}, S., \& {Gotz}, D. 2016, Astronomical and Astrophysical Transactions, 29, 205

\bibitem[{{Eichler} {$et~al$.}(1989){Eichler}, {Livio}, {Piran}, \& {Schramm}}]{eichler89}
{Eichler}, D., {Livio}, M., {Piran}, T., \& {Schramm}, D.~N. 1989, \nat, 340, 126

\bibitem[{{Gill} {$et~al$.}(2020){Gill}, {Granot}, \& {Kumar}}]{Gill_etal_2020}
{Gill}, R., {Granot}, J., \& {Kumar}, P. 2020, \mnras, 491, 3343

\bibitem[{Gill {$et~al$.}(2021)Gill, Kole, \& Granot}]{Gill2021}
Gill, R., Kole, M., \& Granot, J. 2021, Galaxies, 9, doi:10.3390/galaxies9040082

\bibitem[{{G{\"o}tz} {$et~al$.}(2013){G{\"o}tz}, {Covino}, {Fern{\'a}ndez-Soto}, {Laurent}, \& {Bo{\v s}njak}}]{Gotz13}
{G{\"o}tz}, D., {Covino}, S., {Fern{\'a}ndez-Soto}, A., {Laurent}, P., \& {Bo{\v s}njak}, {\v Z}. 2013, MNRAS, 431, 3550

\bibitem[{{G{\"o}tz} {$et~al$.}(2009){G{\"o}tz}, {Laurent}, {Lebrun}, {Daigne}, \& {Bo{\v s}njak}}]{Gotz09}
{G{\"o}tz}, D., {Laurent}, P., {Lebrun}, F., {Daigne}, F., \& {Bo{\v s}njak}, {\v Z}. 2009, Astrophysical Journal Letter, 695, L208

\bibitem[{Gruber {$et~al$.}(2014)Gruber, Goldstein, von Ahlefeld, Bhat, Bissaldi, Briggs, Byrne, Cleveland, Connaughton, Diehl, {$et~al$.}}]{gruber2014fermi}
Gruber, D., Goldstein, A., von Ahlefeld, V.~W., {$et~al$.} 2014, The Astrophysical Journal Supplement Series, 211, 12

\bibitem[{{Gupta} {$et~al$.}(2020{\natexlab{a}}){Gupta}, {Sharma}, {Vibhute}, {Bhattacharya}, {Bhalerao}, {Rao}, {Vadawale}, \& {AstroSat CZTI Collaboration}}]{2020GCN.27680....1G}
{Gupta}, S., {Sharma}, V., {Vibhute}, A., {$et~al$.} 2020{\natexlab{a}}, GRB Coordinates Network, 27680, 1

\bibitem[{{Gupta} {$et~al$.}(2020{\natexlab{b}}){Gupta}, {Vibhute}, {Bhattacharya}, {Bhalerao}, {Rao}, {Vadawale}, \& {AstroSat CZTI Collaboration}}]{2020GCN.28589....1G}
{Gupta}, S., {Vibhute}, V. S.~A., {Bhattacharya}, D., {$et~al$.} 2020{\natexlab{b}}, GRB Coordinates Network, 28589, 1

\bibitem[{Hunter(2007)}]{matplotlib}
Hunter, J.~D. 2007, Computing in Science {\&} Engineering, 9, 90

\bibitem[{{Iwamoto} {$et~al$.}(1998){Iwamoto}, {Mazzali}, {Nomoto}, {Umeda}, {Nakamura}, {Patat}, {Danziger}, {Young}, {Suzuki}, {Shigeyama}, {Augusteijn}, {Doublier}, {Gonzalez}, {Boehnhardt}, {Brewer}, {Hainaut}, {Lidman}, {Leibundgut}, {Cappellaro}, {Turatto}, {Galama}, {Vreeswijk}, {Kouveliotou}, {van Paradijs}, {Pian}, {Palazzi}, \& {Frontera}}]{iwamoto98}
{Iwamoto}, K., {Mazzali}, P.~A., {Nomoto}, K., {$et~al$.} 1998, \nat, 395, 672

\bibitem[{{Iyyani} {$et~al$.}(2015){Iyyani}, {Ryde}, {Ahlgren}, {Burgess}, {Larsson}, {Pe'er}, {Lundman}, {Axelsson}, \& {McGlynn}}]{iyyani15}
{Iyyani}, S., {Ryde}, F., {Ahlgren}, B., {$et~al$.} 2015, \mnras, 450, 1651

\bibitem[{{Klebesadel} {$et~al$.}(1973){Klebesadel}, {Strong}, \& {Olson}}]{discovery}
{Klebesadel}, R.~W., {Strong}, I.~B., \& {Olson}, R.~A. 1973, ApJL, 182, L85

\bibitem[{Kole {$et~al$.}(2020)Kole, {De Angelis}, Berlato, Burgess, Gauvin, Greiner, Hajdas, Li, Li, Pollo, Produit, Rybka, Song, Sun, Szabelski, Tymieniecka, Wang, Wu, Wu, Xiong, Zhang, \& Zhang}]{Kole2020}
Kole, M., {De Angelis}, N., Berlato, F., {$et~al$.} 2020, Astron. Astrophys., 644, doi:10.1051/0004-6361/202037915

\bibitem[{{Kole} {$et~al$.}(2020){Kole}, {De Angelis}, {Berlato}, {Burgess}, {Gauvin}, {Greiner}, {Hajdas}, {Li}, {Li}, {Pollo}, {Produit}, {Rybka}, {Song}, {Sun}, {Szabelski}, {Tymieniecka}, {Wang}, {Wu}, {Wu}, {Xiong}, {Zhang}, \& {Zhang}}]{2020A&A...644A.124K}
{Kole}, M., {De Angelis}, N., {Berlato}, F., {$et~al$.} 2020, \aap, 644, A124

\bibitem[{{Kumar} \& {Zhang}(2015)}]{kumar15}
{Kumar}, P., \& {Zhang}, B. 2015, \physrep, 561, 1

\bibitem[{Landsman(1993)}]{landsman93}
Landsman, W.~B. 1993, Astronomical Data Analysis Software and Systems II, 52

\bibitem[{{Lundman} {$et~al$.}(2018){Lundman}, {Vurm}, \& {Beloborodov}}]{Lundman_etal_2018}
{Lundman}, C., {Vurm}, I., \& {Beloborodov}, A.~M. 2018, \apj, 856, 145

\bibitem[{{MacFadyen} \& {Woosley}(1999)}]{macfadyen99}
{MacFadyen}, A.~I., \& {Woosley}, S.~E. 1999, \apj, 524, 262

\bibitem[{{McConnell}(2017)}]{mcconnell16}
{McConnell}, M.~L. 2017, \nar, 76, 1

\bibitem[{{Narayan} {$et~al$.}(1992){Narayan}, {Paczynski}, \& {Piran}}]{narayan92}
{Narayan}, R., {Paczynski}, B., \& {Piran}, T. 1992, \apjl, 395, L83

\bibitem[{{Piran}(1999)}]{1999PhR...314..575P}
{Piran}, T. 1999, \physrep, 314, 575

\bibitem[{Robitaille {$et~al$.}(2013)Robitaille, Tollerud, Greenfield, Droettboom, Bray, Aldcroft, Davis, Ginsburg, Price-Whelan, Kerzendorf, Conley, Crighton, Barbary, Muna, Ferguson, Grollier, Parikh, Nair, G{\"{u}}nther, Deil, Woillez, Conseil, Kramer, Turner, Singer, Fox, Weaver, Zabalza, Edwards, {Azalee Bostroem}, Burke, Casey, Crawford, Dencheva, Ely, Jenness, Labrie, Lim, Pierfederici, Pontzen, Ptak, Refsdal, Servillat, \& Streicher}]{astropy}
Robitaille, T.~P., Tollerud, E.~J., Greenfield, P., {$et~al$.} 2013, Astronomy {\&} Astrophysics, 558, A33

\bibitem[{{Saraogi} {$et~al$.}(2024){Saraogi}, {Aditya}, {Bhalerao}, {Bala}, {Balasubramanian}, {Mate}, {Chattopadhyay}, {Gupta}, {Prasad}, {Waratkar}, {Navaneeth}, {Gopalakrishnan}, {Bhattacharya}, {Dewangan}, \& {Vadawale}}]{saraogi_2024}
{Saraogi}, D., {Aditya}, J.~V., {Bhalerao}, V., {$et~al$.} 2024, \mnras, 530, 1386

\bibitem[{Scargle(1998)}]{scargle1998studies}
Scargle, J.~D. 1998, The Astrophysical Journal, 504, 405

\bibitem[{{Scargle} {$et~al$.}(2013){Scargle}, {Norris}, {Jackson}, \& {Chiang}}]{Scargle2013}
{Scargle}, J.~D., {Norris}, J.~P., {Jackson}, B., \& {Chiang}, J. 2013, \apj, 764, 167

\bibitem[{{Sharma} {$et~al$.}(2019){Sharma}, {Iyyani}, {Bhattacharya}, {Chattopadhyay}, {Rao}, {Aarthy}, {Vadawale}, {Mithun}, {Bhalerao}, {Ryde}, \& {Pe'er}}]{Sharma_etal_2019}
{Sharma}, V., {Iyyani}, S., {Bhattacharya}, D., {$et~al$.} 2019, \apjl, 882, L10

\bibitem[{{Toma} {$et~al$.}(2009){Toma}, {Sakamoto}, {Zhang}, {Hill}, {McConnell}, {Bloser}, {Yamazaki}, {Ioka}, \& {Nakamura}}]{toma08}
{Toma}, K., {Sakamoto}, T., {Zhang}, B., {$et~al$.} 2009, Astrophysical Journal, 698, 1042

\bibitem[{Toma {$et~al$.}(2009)Toma, Sakamoto, Zhang, Hill, McConnell, Bloser, Yamazaki, Ioka, \& Nakamura}]{toma2009statistical}
Toma, K., Sakamoto, T., Zhang, B., {$et~al$.} 2009, The Astrophysical Journal, 698, 1042

\bibitem[{{Ursi} {$et~al$.}(2020{\natexlab{a}}){Ursi}, {Verrecchia}, {Tavani}, {Casentini}, {Pilia}, {Pittori}, {Argan}, {Cardillo}, {Evangelista}, {Piano}, {Lucarelli}, {Bulgarelli}, {Fioretti}, {Fuschino}, {Parmiggiani}, {Marisaldi}, {Trois}, {Donnarumma}, {Longo}, {Giuliani}, \& {Agile Team}}]{2020GCN.27685....1U}
{Ursi}, A., {Verrecchia}, F., {Tavani}, M., {$et~al$.} 2020{\natexlab{a}}, GRB Coordinates Network, 27685, 1

\bibitem[{{Ursi} {$et~al$.}(2020{\natexlab{b}}){Ursi}, {Verrecchia}, {Pittori}, {Tavani}, {Casentini}, {Argan}, {Cardillo}, {Evangelista}, {Piano}, {Lucarelli}, {Bulgarelli}, {Fioretti}, {Fuschino}, {Parmiggiani}, {Marisaldi}, {Pilia}, {Trois}, {Donnarumma}, {Longo}, {Giuliani}, \& {Agile Team}}]{2020GCN.28588....1U}
{Ursi}, A., {Verrecchia}, F., {Pittori}, C., {$et~al$.} 2020{\natexlab{b}}, GRB Coordinates Network, 28588, 1

\bibitem[{van~der Walt {$et~al$.}(2011)van~der Walt, Colbert, \& Varoquaux}]{numpy}
van~der Walt, S., Colbert, S.~C., \& Varoquaux, G. 2011, Computing in Science {\&} Engineering, 13, 22

\bibitem[{Willis {$et~al$.}(2005)Willis, Barlow, Bird, Clark, Dean, McConnell, Moran, Shaw, \& Sguera}]{Willis2005}
Willis, D.~R., Barlow, E.~J., Bird, A.~J., {$et~al$.} 2005, Astron. Astrophys., 439, 245

\bibitem[{{Woosley}(1993)}]{woosley93}
{Woosley}, S.~E. 1993, \apj, 405, 273

\bibitem[{{Yonetoku} {$et~al$.}(2011){Yonetoku}, {Murakami}, {Gunji}, {Mihara}, {Toma}, {Sakashita}, {Morihara}, {Takahashi}, {Toukairin}, {Fujimoto}, {Kodama}, \& {Kubo}}]{yonetoku11}
{Yonetoku}, D., {Murakami}, T., {Gunji}, S., {$et~al$.} 2011, Astrophysical Journal Letter, 743, L30

\bibitem[{{Yonetoku} {$et~al$.}(2012){Yonetoku}, {Murakami}, {Gunji}, {Mihara}, {Toma}, {Morihara}, {Takahashi}, {Wakashima}, {Yonemochi}, {Sakashita}, {Toukairin}, {Fujimoto}, \& {Kodama}}]{2012ApJ...758L...1Y}
---. 2012, \apjl, 758, L1

\bibitem[{Zhang {$et~al$.}(2016)Zhang, Uhm, Connaughton, Briggs, \& Zhang}]{Zhang2016}
Zhang, B.-B., Uhm, Z.~L., Connaughton, V., Briggs, M.~S., \& Zhang, B. 2016, The Astrophysical Journal, 816, 72

\bibitem[{{Zhang} {$et~al$.}(2019){Zhang}, {Kole}, {Bao}, {Batsch}, {Bernasconi}, {Cadoux}, {Chai}, {Dai}, {Dong}, {Gauvin}, {Hajdas}, {Lan}, {Li}, {Li}, {Li}, {Liu}, {Liu}, {Marcinkowski}, {Produit}, {Orsi}, {Pohl}, {Rybka}, {Shi}, {Song}, {Sun}, {Szabelski}, {Tymieniecka}, {Wang}, {Wang}, {Wen}, {Wu}, {Wu}, {Wu}, {Xiao}, {Xiong}, {Zhang}, {Zhang}, {Zhang}, {Zhang}, \& {Zwolinska}}]{zhang19}
{Zhang}, S.-N., {Kole}, M., {Bao}, T.-W., {$et~al$.} 2019, Nature Astronomy, 3, 258

\end{thebibliography}


\begin{thebibliography}{}
\expandafter\ifx\csname natexlab\endcsname\relax\def\natexlab#1{#1}\fi

\bibitem[{num(2011)}]{numpy}
 2011, Computing in Science {\&} Engineering, 13, 22

\bibitem[{{Agostinelli} {$et~al$.}(2003){Agostinelli}, {Allison}, {Amako},
  {Apostolakis}, {Araujo}, {Arce}, {Asai}, {Axen}, {Banerjee}, {Barrand},
  {Behner}, {Bellagamba}, {Boudreau}, {Broglia}, {Brunengo}, {Burkhardt},
  {Chauvie}, {Chuma}, {Chytracek}, {Cooperman}, {Cosmo}, {Degtyarenko},
  {Dell'Acqua}, {Depaola}, {Dietrich}, {Enami}, {Feliciello}, {Ferguson},
  {Fesefeldt}, {Folger}, {Foppiano}, {Forti}, {Garelli}, {Giani},
  {Giannitrapani}, {Gibin}, {G{\'o}mez Cadenas}, {Gonz{\'a}lez}, {Gracia
  Abril}, {Greeniaus}, {Greiner}, {Grichine}, {Grossheim}, {Guatelli},
  {Gumplinger}, {Hamatsu}, {Hashimoto}, {Hasui}, {Heikkinen}, {Howard},
  {Ivanchenko}, {Johnson}, {Jones}, {Kallenbach}, {Kanaya}, {Kawabata},
  {Kawabata}, {Kawaguti}, {Kelner}, {Kent}, {Kimura}, {Kodama}, {Kokoulin},
  {Kossov}, {Kurashige}, {Lamanna}, {Lamp{\'e}n}, {Lara}, {Lefebure}, {Lei},
  {Liendl}, {Lockman}, {Longo}, {Magni}, {Maire}, {Medernach}, {Minamimoto},
  {Mora de Freitas}, {Morita}, {Murakami}, {Nagamatu}, {Nartallo}, {Nieminen},
  {Nishimura}, {Ohtsubo}, {Okamura}, {O'Neale}, {Oohata}, {Paech}, {Perl},
  {Pfeiffer}, {Pia}, {Ranjard}, {Rybin}, {Sadilov}, {Di Salvo}, {Santin},
  {Sasaki}, {Savvas}, {Sawada}, {Scherer}, {Sei}, {Sirotenko}, {Smith},
  {Starkov}, {Stoecker}, {Sulkimo}, {Takahata}, {Tanaka}, {Tcherniaev}, {Safai
  Tehrani}, {Tropeano}, {Truscott}, {Uno}, {Urban}, {Urban}, {Verderi},
  {Walkden}, {Wander}, {Weber}, {Wellisch}, {Wenaus}, {Williams}, {Wright},
  {Yamada}, {Yoshida}, {Zschiesche}, \& {G EANT4
  Collaboration}}]{2003NIMPA.506..250A}
{Agostinelli}, S., {Allison}, J., {Amako}, K., {$et~al$.} 2003, Nuclear
  Instruments and Methods in Physics Research A, 506, 250

\bibitem[{{Allison} {$et~al$.}(2006){Allison}, {Amako}, {Apostolakis},
  {Araujo}, {Arce Dubois}, {Asai}, {Barrand}, {Capra}, {Chauvie}, {Chytracek},
  {Cirrone}, {Cooperman}, {Cosmo}, {Cuttone}, {Daquino}, {Donszelmann},
  {Dressel}, {Folger}, {Foppiano}, {Generowicz}, {Grichine}, {Guatelli},
  {Gumplinger}, {Heikkinen}, {Hrivnacova}, {Howard}, {Incerti}, {Ivanchenko},
  {Johnson}, {Jones}, {Koi}, {Kokoulin}, {Kossov}, {Kurashige}, {Lara},
  {Larsson}, {Lei}, {Link}, {Longo}, {Maire}, {Mantero}, {Mascialino},
  {McLaren}, {Mendez Lorenzo}, {Minamimoto}, {Murakami}, {Nieminen}, {Pandola},
  {Parlati}, {Peralta}, {Perl}, {Pfeiffer}, {Pia}, {Ribon}, {Rodrigues},
  {Russo}, {Sadilov}, {Santin}, {Sasaki}, {Smith}, {Starkov}, {Tanaka},
  {Tcherniaev}, {Tome}, {Trindade}, {Truscott}, {Urban}, {Verderi}, {Walkden},
  {Wellisch}, {Williams}, {Wright}, \& {Yoshida}}]{2006ITNS...53..270A}
{Allison}, J., {Amako}, K., {Apostolakis}, J., {$et~al$.} 2006, IEEE
  Transactions on Nuclear Science, 53, 270

\bibitem[{{Allison} {$et~al$.}(2016){Allison}, {Amako}, {Apostolakis}, {Arce},
  {Asai}, {Aso}, {Bagli}, {Bagulya}, {Banerjee}, {Barrand}, {Beck}, {Bogdanov},
  {Brandt}, {Brown}, {Burkhardt}, {Canal}, {Cano-Ott}, {Chauvie}, {Cho},
  {Cirrone}, {Cooperman}, {Cort{\'e}s-Giraldo}, {Cosmo}, {Cuttone}, {Depaola},
  {Desorgher}, {Dong}, {Dotti}, {Elvira}, {Folger}, {Francis}, {Galoyan},
  {Garnier}, {Gayer}, {Genser}, {Grichine}, {Guatelli}, {Gu{\`e}ye},
  {Gumplinger}, {Howard}, {H{\v r}ivn{\'a}{\v c}ov{\'a}}, {Hwang}, {Incerti},
  {Ivanchenko}, {Ivanchenko}, {Jones}, {Jun}, {Kaitaniemi}, {Karakatsanis},
  {Karamitrosi}, {Kelsey}, {Kimura}, {Koi}, {Kurashige}, {Lechner}, {Lee},
  {Longo}, {Maire}, {Mancusi}, {Mantero}, {Mendoza}, {Morgan}, {Murakami},
  {Nikitina}, {Pandola}, {Paprocki}, {Perl}, {Petrovi{\'c}}, {Pia}, {Pokorski},
  {Quesada}, {Raine}, {Reis}, {Ribon}, {Risti{\'c} Fira}, {Romano}, {Russo},
  {Santin}, {Sasaki}, {Sawkey}, {Shin}, {Strakovsky}, {Taborda}, {Tanaka},
  {Tom{\'e}}, {Toshito}, {Tran}, {Truscott}, {Urban}, {Uzhinsky}, {Verbeke},
  {Verderi}, {Wendt}, {Wenzel}, {Wright}, {Wright}, {Yamashita}, {Yarba}, \&
  {Yoshida}}]{2016NIMPA.835..186A}
---. 2016, Nuclear Instruments and Methods in Physics Research A, 835, 186

\bibitem[{{Antia} {$et~al$.}(2013){Antia}, {Chitinus}, {Katoch}, {Mazumdar},
  {Pahari}, {Vadawale}, \& {Yadav}}]{laxpcMM}
{Antia}, H.~M., {Chitinus}, V.~R., {Katoch}, T.~V., {$et~al$.} 2013, {GEANT4
  simulation of LAXPC background}, Tech. rep., {Tata Institute of Fundamental
  Research (TIFR)}

\bibitem[{{Band} {$et~al$.}(1993){Band}, {Matteson}, {Ford}, {Schaefer},
  {Palmer}, {Teegarden}, {Cline}, {Briggs}, {Paciesas}, {Pendleton}, {Fishman},
  {Kouveliotou}, {Meegan}, {Wilson}, \& {Lestrade}}]{1993ApJ...413..281B}
{Band}, D., {Matteson}, J., {Ford}, L., {$et~al$.} 1993, \apj, 413, 281

\bibitem[{{Bhalerao} {$et~al$.}(2015){Bhalerao}, {Bhattacharya}, {Rao}, \&
  {Vadawale}}]{2015GCN..18422...1B}
{Bhalerao}, V., {Bhattacharya}, D., {Rao}, A.~R., \& {Vadawale}, S. 2015, GRB
  Coordinates Network, 18422

\bibitem[{{Bhalerao} {$et~al$.}(2017){Bhalerao}, {Bhattacharya}, {Rao}, \&
  {Vadawale}}]{2017GCN..20412...1B}
---. 2017, GRB Coordinates Network, 20412

\bibitem[{Bhalerao {$et~al$.}(2017{\natexlab{a}})Bhalerao, Kasliwal,
  Bhattacharya, Corsi, Aarthy, Adams, Blagorodnova, Cantwell, Cenko, Fender,
  Frail, Itoh, Jencson, Kawai, Kong, Kupfer, Kutyrev, Mao, Mate, Mithun,
  Mooley, Perley, Perrott, Quimby, Rao, Singer, Sharma, Titterington, Troja,
  Vadawale, Vibhute, Vedantham, \& Veilleux}]{bkb+17}
Bhalerao, V., Kasliwal, M., Bhattacharya, D., {$et~al$.} 2017{\natexlab{a}},
  Astrophysical Journal, 845, doi:10.3847/1538-4357/aa81d2

\bibitem[{Bhalerao {$et~al$.}(2017{\natexlab{b}})Bhalerao, Bhattacharya,
  Vibhute, Pawar, Rao, Hingar, Khanna, Kutty, Malkar, Patil, Arora, Sinha,
  Priya, Samuel, Sreekumar, Vinod, Mithun, Vadawale, Vagshette, Navalgund,
  Sarma, Pandiyan, Seetha, \& Subbarao}]{czti}
Bhalerao, V., Bhattacharya, D., Vibhute, A., {$et~al$.} 2017{\natexlab{b}},
  Journal of Astrophysics and Astronomy, 38, 31

\bibitem[{{Bhattacharya} {$et~al$.}(2016){Bhattacharya}, {Dewangan},
  {Pandiyan}, {Ramadevi}, {Rao}, {Singh}, {Tandon}, \& {Yadav}}]{astrosatHB}
{Bhattacharya}, D., {Dewangan}, G.~C., {Pandiyan}, R., {$et~al$.} 2016,
  {AstroSat Handbook}, Tech. rep., {Space Science Programme Office, INDIAN
  SPACE RESEARCH ORGANISATION (ISRO)}

\bibitem[{{Bissaldi} {$et~al$.}(2018){Bissaldi}, {Kocevski}, \&
  {Omodei}}]{GRB180914A_loc}
{Bissaldi}, E., {Kocevski}, D., \& {Omodei}, N. 2018, GRB Coordinates Network,
  23225, 1

\bibitem[{{Bissaldi} \& {Meegan}(2019)}]{GRB190530A_b}
{Bissaldi}, E., \& {Meegan}, C. 2019, GRB Coordinates Network, 24692, 1

\bibitem[{Blackburn(1995)}]{blackburn95}
Blackburn, J.~K. 1995, Astronomical Data Analysis Software and Systems IV, 77

\bibitem[{Chand {$et~al$.}(2019)Chand, Chattopadhyay, Oganesyan, Rao, Vadawale,
  Bhattacharya, Bhalerao, \& Misra}]{2019ApJ...874...70C}
Chand, V., Chattopadhyay, T., Oganesyan, G., {$et~al$.} 2019, $\backslash$apj,
  874, 70

\bibitem[{{Chattopadhyay} {$et~al$.}(2014){Chattopadhyay}, {Vadawale}, {Rao},
  {Sreekumar}, \& {Bhattacharya}}]{2014ExA....37..555C}
{Chattopadhyay}, T., {Vadawale}, S.~V., {Rao}, A.~R., {Sreekumar}, S., \&
  {Bhattacharya}, D. 2014, Experimental Astronomy, 37, 555

\bibitem[{Chattopadhyay {$et~al$.}(2019)Chattopadhyay, Vadawale, Aarthy,
  Mithun, Chand, Ratheesh, Basak, Rao, Bhalerao, Mate, Arvind, Sharma, \&
  Bhattacharya}]{2019ApJ...884..123C}
Chattopadhyay, T., Vadawale, S.~V., Aarthy, E., {$et~al$.} 2019,
  $\backslash$apj, 884, 123

\bibitem[{{D'Avanzo} {$et~al$.}(2017){D'Avanzo}, {Evans}, {Kuin}, {Page},
  {Palmer}, \& {Sbarufatti}}]{GRB171027A_loc}
{D'Avanzo}, P., {Evans}, P.~A., {Kuin}, N.~P.~M., {$et~al$.} 2017, GRB
  Coordinates Network, 22053, 1

\bibitem[{{Frederiks} {$et~al$.}(2017{\natexlab{a}}){Frederiks}, {Golenetskii},
  {Aptekar}, {Oleynik}, {Ulanov}, {Svinkin}, {Tsvetkova}, {Lysenko}, {Kozlova},
  \& {Cline}}]{GRB170115B_a}
{Frederiks}, D., {Golenetskii}, S., {Aptekar}, R., {$et~al$.}
  2017{\natexlab{a}}, GRB Coordinates Network, 20476, 1

\bibitem[{{Frederiks} {$et~al$.}(2017{\natexlab{b}}){Frederiks}, {Golenetskii},
  {Aptekar}, {Oleynik}, {Ulanov}, {Svinkin}, {Tsvetkova}, {Lysenko}, {Kozlova},
  \& {Cline}}]{GRB170527A_a}
---. 2017{\natexlab{b}}, GRB Coordinates Network, 21166, 1

\bibitem[{{Frederiks} {$et~al$.}(2017{\natexlab{c}}){Frederiks}, {Golenetskii},
  {Aptekar}, {Oleynik}, {Ulanov}, {Svinkin}, {Tsvetkova}, {Lysenko}, {Kozlova},
  \& {Cline}}]{GRB171010A_a}
---. 2017{\natexlab{c}}, GRB Coordinates Network, 22003, 1

\bibitem[{{Frederiks} {$et~al$.}(2017{\natexlab{d}}){Frederiks}, {Golenetskii},
  {Aptekar}, {Oleynik}, {Ulanov}, {Svinkin}, {Tsvetkova}, {Lysenko}, {Kozlova},
  \& {Cline}}]{GRB171027A_a}
---. 2017{\natexlab{d}}, GRB Coordinates Network, 22070, 1

\bibitem[{{Frederiks} {$et~al$.}(2018{\natexlab{a}}){Frederiks}, {Golenetskii},
  {Aptekar}, {Ulanov}, {Svinkin}, {Tsvetkova}, {Lysenko}, {Kozlova}, \&
  {Cline}}]{GRB180325A_a}
---. 2018{\natexlab{a}}, GRB Coordinates Network, 22546, 1

\bibitem[{{Frederiks} {$et~al$.}(2018{\natexlab{b}}){Frederiks}, {Golenetskii},
  {Aptekar}, {Kozlova}, {Lysenko}, {Svinkin}, {Tsvetkova}, {Ulanov}, \&
  {Cline}}]{GRB180728A_a}
---. 2018{\natexlab{b}}, GRB Coordinates Network, 23061, 1

\bibitem[{{Frederiks} {$et~al$.}(2019{\natexlab{a}}){Frederiks}, {Golenetskii},
  {Aptekar}, {Kozlova}, {Lysenko}, {Svinkin}, {Tsvetkova}, {Ulanov}, {Cline},
  \& {Konus-Wind Team}}]{GRB190530A_a}
---. 2019{\natexlab{a}}, GRB Coordinates Network, 24715, 1

\bibitem[{{Frederiks} {$et~al$.}(2019{\natexlab{b}}){Frederiks}, {Golenetskii},
  {Aptekar}, {Kozlova}, {Lysenko}, {Svinkin}, {Tsvetkova}, {Ulanov}, \&
  {Cline}}]{GRB190117A_a}
---. 2019{\natexlab{b}}, GRB Coordinates Network, 23782, 1

\bibitem[{Hunter(2007)}]{matplotlib}
Hunter, J.~D. 2007, Computing in Science {\&} Engineering, 9, 90

\bibitem[{{Hurley} {$et~al$.}(2017){Hurley}, {Mitrofanov}, {Golovin}, {Litvak},
  {Sanin}, {Svinkin}, {Golenetskii}, {Aptekar}, {Frederiks}, {Kozlova},
  {Cline}, {Connaughton}, {Briggs}, {Meegan}, {Pelassa}, {Goldstein}, {von
  Kienlin}, {Zhang}, {Rau}, {Savchenko}, {Bozzo}, {Ferrigno}, {Barthelmy},
  {Cummings}, {Krimm}, {Palmer}, {Boynton}, {Fellows}, {Harshman}, {Enos}, \&
  {Starr}}]{GRB170527A_loc}
{Hurley}, K., {Mitrofanov}, I.~G., {Golovin}, D., {$et~al$.} 2017, GRB
  Coordinates Network, 21164, 1

\bibitem[{{Hurley} {$et~al$.}(2018){Hurley}, {Mitrofanov}, {Golovin}, {Litvak},
  {Sanin}, {Svinkin}, {Golenetskii}, {Aptekar}, {Frederiks}, {Kozlova},
  {Cline}, {Connaughton}, {Briggs}, {Meegan}, {Pelassa}, {Goldstein}, {von
  Kienlin}, {Zhang}, {Rau}, {Savchenko}, {Bozzo}, {Ferrigno}, {Barthelmy},
  {Cummings}, {Krimm}, {Palmer}, {Boynton}, {Fellows}, {Harshman}, {Enos}, \&
  {Starr}}]{GRB180427A_loc}
---. 2018, GRB Coordinates Network, 22679, 1

\bibitem[{{Hurley} {$et~al$.}(2019){Hurley}, {Mitrofanov}, {Golovin}, {Litvak},
  {Sanin}, {Svinkin}, {Golenetskii}, {Aptekar}, {Frederiks}, {Kozlova},
  {Cline}, {von Kienlin}, {Zhang}, {Rau}, {Savchenko}, {Bozzo}, {Ferrigno},
  {Barthelmy}, {Cummings}, {Krimm}, {Palmer}, {Xiao}, {Li}, {Li}, {Huang},
  {Xiong}, {Boynton}, {Fellows}, {Harshman}, {Enos}, \&
  {Starr}}]{GRB190117A_loc}
---. 2019, GRB Coordinates Network, 23764, 1

\bibitem[{Kasliwal {$et~al$.}(2017)Kasliwal, Nakar, Singer, Kaplan, Cook, {Van
  Sistine}, Lau, Fremling, Gottlieb, Jencson, Adams, Feindt, Hotokezaka, Ghosh,
  Perley, Yu, Piran, Allison, Anupama, Balasubramanian, Bannister, Bally,
  Barnes, Barway, Bellm, Bhalerao, Bhattacharya, Blagorodnova, Bloom, Brady,
  Cannella, Chatterjee, Cenko, Cobb, Copperwheat, Corsi, De, Dobie, Emery,
  Evans, Fox, Frail, Frohmaier, Goobar, Hallinan, Harrison, Helou, Hinderer,
  Ho, Horesh, Ip, Itoh, Kasen, Kim, Kuin, Kupfer, Lynch, Madsen, Mazzali,
  Miller, Mooley, Murphy, Ngeow, Nichols, Nissanke, Nugent, Ofek, Qi, Quimby,
  Rosswog, Rusu, Sadler, Schmidt, Sollerman, Steele, Williamson, Xu, Yan,
  Yatsu, Zhang, \& Zhao}]{kns+17}
Kasliwal, M., Nakar, E., Singer, L., {$et~al$.} 2017, Science, 358, 1559

\bibitem[{{Kozlova} {$et~al$.}(2016){Kozlova}, {Golenetskii}, {Aptekar},
  {Frederiks}, {Oleynik}, {Ulanov}, {Svinkin}, {Tsvetkova}, {Lysenko}, \&
  {Cline}}]{GRB160821A_a}
{Kozlova}, A., {Golenetskii}, S., {Aptekar}, R., {$et~al$.} 2016, GRB
  Coordinates Network, 19842, 1

\bibitem[{{Kozlova} {$et~al$.}(2017){Kozlova}, {Golenetskii}, {Aptekar},
  {Frederiks}, {Oleynik}, {Ulanov}, {Svinkin}, {Tsvetkova}, {Lysenko}, \&
  {Cline}}]{GRB170921B_a}
---. 2017, GRB Coordinates Network, 21926, 1

\bibitem[{{Kozlova} {$et~al$.}(2018){Kozlova}, {Golenetskii}, {Aptekar},
  {Frederiks}, {Ulanov}, {Svinkin}, {Tsvetkova}, {Lysenko}, \&
  {Cline}}]{GRB180427A_a}
---. 2018, GRB Coordinates Network, 22680, 1

\bibitem[{Landsman(1993)}]{landsman93}
Landsman, W.~B. 1993, Astronomical Data Analysis Software and Systems II, 52

\bibitem[{{Marcinkowski} {$et~al$.}(2017){Marcinkowski}, {Xiao}, \&
  {Hajdas}}]{2017GCN..20387...1M}
{Marcinkowski}, R., {Xiao}, H., \& {Hajdas}, W. 2017, GRB Coordinates Network,
  20387

\bibitem[{{Melandri} {$et~al$.}(2017){Melandri}, {D'Avanzo}, {Page},
  {Sakamoto}, {Sbarufatti}, {Starling}, \& {Troja}}]{GRB170822A_loc}
{Melandri}, A., {D'Avanzo}, P., {Page}, K.~L., {$et~al$.} 2017, GRB Coordinates
  Network, 21640, 1

\bibitem[{{Moss} {$et~al$.}(2018){Moss}, {Barthelmy}, {D'Elia}, {Gronwall},
  {Gropp}, {Kennea}, {Lien}, {Palmer}, \& {Tohuvavohu}}]{GRB180809B_loc}
{Moss}, M.~J., {Barthelmy}, S.~D., {D'Elia}, V., {$et~al$.} 2018, GRB
  Coordinates Network, 23105, 1

\bibitem[{{Poole} {$et~al$.}(2011){Poole}, {Cornelius}, {Trapp}, \&
  {Langton}}]{2011arXiv1105.0963P}
{Poole}, C.~M., {Cornelius}, I., {Trapp}, J.~V., \& {Langton}, C.~M. 2011,
  ArXiv e-prints

\bibitem[{Rao {$et~al$.}(2017)Rao, Bhattacharya, Bhalerao, Vadawale, \&
  Sreekumar}]{Rao2017a}
Rao, A., Bhattacharya, D., Bhalerao, V., Vadawale, S., \& Sreekumar, S. 2017,
  Current Science, 113, doi:10.18520/cs/v113/i04/595-598

\bibitem[{{Rao} {$et~al$.}(2016){Rao}, {Chand}, {Hingar}, {Iyyani}, {Khanna},
  {Kutty}, {Malkar}, {Paul}, {Bhalerao}, {Bhattacharya}, {Dewangan}, {Pawar},
  {Vibhute}, {Chattopadhyay}, {Mithun}, {Vadawale}, {Vagshette}, {Basak},
  {Pradeep}, {Samuel}, {Sreekumar}, {Vinod}, {Navalgund}, {Pandiyan}, {Sarma},
  {Seetha}, \& {Subbarao}}]{2016ApJ...833...86R}
{Rao}, A.~R., {Chand}, V., {Hingar}, M.~K., {$et~al$.} 2016, \apj, 833, 86

\bibitem[{Robitaille {$et~al$.}(2013)Robitaille, Tollerud, Greenfield,
  Droettboom, Bray, Aldcroft, Davis, Ginsburg, Price-Whelan, Kerzendorf,
  Conley, Crighton, Barbary, Muna, Ferguson, Grollier, Parikh, Nair,
  G{\"{u}}nther, Deil, Woillez, Conseil, Kramer, Turner, Singer, Fox, Weaver,
  Zabalza, Edwards, {Azalee Bostroem}, Burke, Casey, Crawford, Dencheva, Ely,
  Jenness, Labrie, Lim, Pierfederici, Pontzen, Ptak, Refsdal, Servillat, \&
  Streicher}]{astropy}
Robitaille, T.~P., Tollerud, E.~J., Greenfield, P., {$et~al$.} 2013, Astronomy
  {\&} Astrophysics, 558, A33

\bibitem[{{Sbarufatti} {$et~al$.}(2017){Sbarufatti}, {Burrows}, {Beardmore},
  {Gibson}, {D'Ai}, {Melandri}, {D'Avanzo}, {Cholden-Brown}, {Evans}, \&
  {Breeveld}}]{GRB170121B_loc}
{Sbarufatti}, B., {Burrows}, D.~N., {Beardmore}, A.~P., {$et~al$.} 2017, GRB
  Coordinates Network, 20510, 1

\bibitem[{{Sharma} {$et~al$.}(2017){Sharma}, {Bhalerao}, {Bhattacharya}, {Rao},
  \& {Vadawale}}]{2017GCN..20389...1S}
{Sharma}, V., {Bhalerao}, V., {Bhattacharya}, D., {Rao}, A.~R., \& {Vadawale},
  S. 2017, GRB Coordinates Network, 20389

\bibitem[{Sharma {$et~al$.}(2020)Sharma, Marathe, Bhalerao, Shenoy, Waratkar,
  Nadella, Page, Hebbar, Vibhute, Bhattacharya, Rao, \&
  Vadawale}]{sharma2020search}
Sharma, Y., Marathe, A., Bhalerao, V., {$et~al$.} 2020, The Search for Fast
  Transients with CZTI, arXiv:2011.07067

\bibitem[{{Singh} {$et~al$.}(2014){Singh}, {Tandon}, {Agrawal}, {Antia},
  {Manchanda}, {Yadav}, {Seetha}, {Ramadevi}, {Rao}, {Bhattacharya}, {Paul},
  {Sreekumar}, {Bhattacharyya}, {Stewart}, {Hutchings}, {Annapurni}, {Ghosh},
  {Murthy}, {Pati}, {Rao}, {Stalin}, {Girish}, {Sankarasubramanian},
  {Vadawale}, {Bhalerao}, {Dewangan}, {Dedhia}, {Hingar}, {Katoch}, {Kothare},
  {Mirza}, {Mukerjee}, {Shah}, {Shah}, {Mohan}, {Sangal}, {Nagabhusana},
  {Sriram}, {Malkar}, {Sreekumar}, {Abbey}, {Hansford}, {Beardmore}, {Sharma},
  {Murthy}, {Kulkarni}, {Meena}, {Babu}, \& {Postma}}]{2014SPIE.9144E..1SS}
{Singh}, K.~P., {Tandon}, S.~N., {Agrawal}, P.~C., {$et~al$.} 2014, in
  \procspie, Vol. 9144, Space Telescopes and Instrumentation 2014: Ultraviolet
  to Gamma Ray, 91441S

\bibitem[{{Svinkin} {$et~al$.}(2016{\natexlab{a}}){Svinkin}, {Golenetskii},
  {Aptekar}, {Frederiks}, {Kozlova}, {Cline}, {Hurley}, {von Kienlin}, {Zhang},
  {Rau}, {Savchenko}, {Bozzo}, \& {Ferrigno}}]{GRB160530A_loc}
{Svinkin}, D., {Golenetskii}, S., {Aptekar}, R., {$et~al$.} 2016{\natexlab{a}},
  GRB Coordinates Network, 19476, 1

\bibitem[{{Svinkin} {$et~al$.}(2016{\natexlab{b}}){Svinkin}, {Golenetskii},
  {Aptekar}, {Frederiks}, {Oleynik}, {Ulanov}, {Tsvetkova}, {Lysenko},
  {Kozlova}, \& {Cline}}]{GRB160530A_a}
---. 2016{\natexlab{b}}, GRB Coordinates Network, 19477, 1

\bibitem[{{Svinkin} {$et~al$.}(2016{\natexlab{c}}){Svinkin}, {Golenetskii},
  {Aptekar}, {Frederiks}, {Oleynik}, {Ulanov}, {Tsvetkova}, {Lysenko},
  {Kozlova}, \& {Cline}}]{GRB160720_a}
---. 2016{\natexlab{c}}, GRB Coordinates Network, 19727, 1

\bibitem[{{Svinkin} {$et~al$.}(2017){Svinkin}, {Golenetskii}, {Aptekar},
  {Frederiks}, {Oleynik}, {Ulanov}, {Tsvetkova}, {Lysenko}, {Kozlova}, \&
  {Cline}}]{GRB170822A_a}
---. 2017, GRB Coordinates Network, 21679, 1

\bibitem[{{Svinkin} {$et~al$.}(2018){Svinkin}, {Golenetskii}, {Aptekar},
  {Frederiks}, {Ulanov}, {Tsvetkova}, {Lysenko}, {Kozlova}, \&
  {Cline}}]{GRB180809B_a}
---. 2018, GRB Coordinates Network, 23128, 1

\bibitem[{{Troja} {$et~al$.}(2018){Troja}, {D'Ai}, {D'Elia}, {Kennea},
  {Laporte}, {Lien}, {Marshall}, {Palmer}, {Siegel}, \&
  {Tohuvavohu}}]{GRB180325A_loc}
{Troja}, E., {D'Ai}, A., {D'Elia}, V., {$et~al$.} 2018, GRB Coordinates
  Network, 22532, 1

\bibitem[{{Tsvetkova} {$et~al$.}(2016{\natexlab{a}}){Tsvetkova}, {Golenetskii},
  {Aptekar}, {Frederiks}, {Oleynik}, {Ulanov}, {Svinkin}, {Lysenko}, {Kozlova},
  \& {Cline}}]{GRB160325A_a}
{Tsvetkova}, A., {Golenetskii}, S., {Aptekar}, R., {$et~al$.}
  2016{\natexlab{a}}, GRB Coordinates Network, 19244, 1

\bibitem[{{Tsvetkova} {$et~al$.}(2016{\natexlab{b}}){Tsvetkova}, {Golenetskii},
  {Aptekar}, {Frederiks}, {Oleynik}, {Ulanov}, {Svinkin}, {Lysenko}, {Kozlova},
  \& {Cline}}]{GRB160607A_a}
---. 2016{\natexlab{b}}, GRB Coordinates Network, 19511, 1

\bibitem[{{Tsvetkova} {$et~al$.}(2017){Tsvetkova}, {Golenetskii}, {Aptekar},
  {Frederiks}, {Oleynik}, {Ulanov}, {Svinkin}, {Lysenko}, {Kozlova}, \&
  {Cline}}]{GRB170228A_a}
---. 2017, GRB Coordinates Network, 20806, 1

\bibitem[{{Tsvetkova} {$et~al$.}(2018{\natexlab{a}}){Tsvetkova}, {Golenetskii},
  {Aptekar}, {Frederiks}, {Oleynik}, {Ulanov}, {Svinkin}, {Lysenko}, {Kozlova},
  \& {Cline}}]{GRB180314A_a}
---. 2018{\natexlab{a}}, GRB Coordinates Network, 22513, 1

\bibitem[{{Tsvetkova} {$et~al$.}(2018{\natexlab{b}}){Tsvetkova}, {Golenetskii},
  {Aptekar}, {Frederiks}, {Ulanov}, {Svinkin}, {Lysenko}, {Kozlova}, \&
  {Cline}}]{GRB180914A_a}
---. 2018{\natexlab{b}}, GRB Coordinates Network, 23254, 1

\bibitem[{{Tsvetkova} {$et~al$.}(2019){Tsvetkova}, {Golenetskii}, {Aptekar},
  {Frederiks}, {Ulanov}, {Svinkin}, {Lysenko}, {Kozlova}, \&
  {Cline}}]{GRB190519A_a}
---. 2019, GRB Coordinates Network, 24652, 1

\bibitem[{{Ukwatta} {$et~al$.}(2016){Ukwatta}, {Beardmore}, {Evans}, {Gehrels},
  {Kennea}, {Krimm}, {Page}, \& {Palmer}}]{GRB160607_loc}
{Ukwatta}, T.~N., {Beardmore}, A.~P., {Evans}, P.~A., {$et~al$.} 2016, GRB
  Coordinates Network, 19502, 1

\bibitem[{Vadawale {$et~al$.}(2018)Vadawale, Chattopadhyay, Mithun, Rao,
  Bhattacharya, Vibhute, Bhalerao, Dewangan, Misra, Paul, Basu, Joshi,
  Sreekumar, Samuel, Priya, Vinod, \& Seetha}]{2018NatAs...2...50V}
Vadawale, S., Chattopadhyay, T., Mithun, N., {$et~al$.} 2018, Nature Astronomy,
  2, 50

\bibitem[{{Vadawale} {$et~al$.}(2015){Vadawale}, {Chattopadhyay}, {Rao},
  {Bhattacharya}, {Bhalerao}, {Vagshette}, {Pawar}, \&
  {Sreekumar}}]{2015A&A...578A..73V}
{Vadawale}, S.~V., {Chattopadhyay}, T., {Rao}, A.~R., {$et~al$.} 2015, \aap,
  578, A73

\bibitem[{{Veres} {$et~al$.}(2018){Veres}, {Meegan}, \&
  {Mailyan}}]{GRB180728A_b}
{Veres}, P., {Meegan}, C., \& {Mailyan}, B. 2018, GRB Coordinates Network,
  23053, 1

\bibitem[{{von Kienlin}(2019)}]{GRB190519A_b}
{von Kienlin}, A. 2019, GRB Coordinates Network, 24596, 1

\bibitem[{{von Kienlin} {$et~al$.}(2020){von Kienlin}, {Meegan}, {Paciesas},
  {Bhat}, {Bissaldi}, {Briggs}, {Burns}, {Cleveland}, {Gibby}, {Giles},
  {Goldstein}, {Hamburg}, {Hui}, {Kocevski}, {Mailyan}, {Malacaria},
  {Poolakkil}, {Preece}, {Roberts}, {Veres}, \&
  {Wilson-Hodge}}]{vonKienlin2020}
{von Kienlin}, A., {Meegan}, C.~A., {Paciesas}, W.~S., {$et~al$.} 2020, \apj,
  893, 46

\bibitem[{{Wanderman} \& {Piran}(2010)}]{2010MNRAS.406.1944W}
{Wanderman}, D., \& {Piran}, T. 2010, \mnras, 406, 1944

\bibitem[{{Wanderman} \& {Piran}(2015)}]{2015MNRAS.448.3026W}
---. 2015, \mnras, 448, 3026

\end{thebibliography}


\appendix


\end{document}